\documentclass[sigconf]{acmart}
\AtBeginDocument{%
  }

\copyrightyear{2025}
\acmYear{2025}
\setcopyright{acmlicensed}
\acmConference[SIGMOD-Companion'25]{Companion of the 2025 International Conference on Management of
Data}{June 22--27, 2025}{Berlin, Germany}
\acmBooktitle{Companion of the 2025 International Conference on
Management of Data (SIGMOD-Companion '25), June 22--27, 2025, Berlin,
Germany}
\acmDOI{10.1145/3722212.3724426}
\acmISBN{979-8-4007-1564-8/2025/06}

\usepackage{booktabs}   
\usepackage{subcaption} 
\usepackage{algorithm}
\usepackage{algpseudocode}
\usepackage{url}      
\usepackage{hyperref} 
\usepackage{booktabs} 
\usepackage{tabularx} 
\usepackage{xcolor} 
\usepackage{caption}
\usepackage{marvosym}
\usepackage{balance}

\begin{document}

\title[ABase]{
    ABase: the Multi-Tenant NoSQL Serverless Database for Diverse and Dynamic Workloads in Large-scale Cloud Environments
}

\author{Rong Kang}
\orcid{0009-0005-8449-0223}
\affiliation{%
  \institution{ByteDance Inc.}
  \city{Beijing}
  \country{China}
}
\email{kangrong.cn@bytedance.com}

\author{Yanbin Chen}
\orcid{0009-0002-2960-6914}
\affiliation{%
  \institution{ByteDance Inc.}
  \city{Beijing}
  \country{China}
}
\email{chenyanbin.cyb@bytedance.com}

\author{Ye Liu}
\orcid{0009-0009-0255-4481}
\affiliation{%
  \institution{Bytedance Inc.}
  \city{San Jose}
  \country{USA}
}
\email{ye.liu@bytedance.com}

\author{Fuxin Jiang}
\orcid{0000-0002-7522-061X}
\affiliation{%
  \institution{ByteDance Inc.}
  \city{Beijing}
  \country{China}
}
\email{jiangfuxin@bytedance.com}

\author{Qingshuo Li}
\orcid{0009-0006-2372-6797}
\affiliation{%
  \institution{ByteDance Inc.}
  \city{Beijing}
  \country{China}
}
\email{liqingshuo.liqs@bytedance.com}

\author{Miao Ma}
\orcid{0009-0005-4274-5849}
\affiliation{%
  \institution{ByteDance Inc.}
  \city{Sydney}
  \country{Australia}
}
\email{miao.ma@bytedance.com}

\author{Jian Liu}
\orcid{0009-0008-0260-8833}
\affiliation{%
  \institution{ByteDance Inc.}
  \city{Beijing}
  \country{China}
}
\email{liujian.kv@bytedance.com}
\authornote{Jian Liu is the corresponding author.}

\author{Guangliang Zhao}
\orcid{0009-0004-8853-3588}
\affiliation{%
  \institution{ByteDance Inc.}
  \city{Hangzhou}
  \country{China}
}
\email{gaoliang6@bytedance.com}

\author{Tieying Zhang}
\orcid{0009-0003-2250-5528}
\affiliation{%
  \institution{Bytedance Inc.}
  \city{San Jose}
  \country{USA}
}
\email{tieying.zhang@bytedance.com}

\author{Jianjun Chen}
\orcid{0000-0002-3734-892X}
\affiliation{%
  \institution{Bytedance Inc.}
  \city{San Jose}
  \country{USA}
}
\email{jianjun.chen@bytedance.com}

\author{Lei Zhang}
\orcid{0009-0004-1681-1956}
\affiliation{%
  \institution{ByteDance Inc.}
  \city{Chengdu}
  \country{China}
}
\email{zhanglei.michael@bytedance.com}

\renewcommand{\shortauthors}{Rong Kang et al.}

\begin{abstract}
Multi-tenant architectures enhance the elasticity and resource utilization of NoSQL databases by allowing multiple tenants to co-locate and share resources. However, in large-scale cloud environments, the diverse and dynamic nature of workloads poses significant challenges for multi-tenant NoSQL databases. Based on our practical observations, we have identified three crucial challenges: (1) the impact of caching on performance isolation, as cache hits alter request execution and resource consumption, leading to inaccurate traffic control; (2) the dynamic changes in traffic, with changes in tenant traffic trends causing throttling or resource wastage, and changes in access distribution causing hot key pressure or cache hit ratio drops; and (3) the imbalanced layout of data nodes due to tenants' diverse resource requirements, leading to low resource utilization. To address these challenges, we introduce ABase, a multi-tenant NoSQL serverless database developed at ByteDance. ABase introduces a two-layer caching mechanism with a cache-aware isolation mechanism to ensure accurate resource consumption estimates. Furthermore, ABase employs a predictive autoscaling policy to dynamically adjust resources in response to tenant traffic changes and a multi-resource rescheduling algorithm to balance resource utilization across data nodes. With these innovations, ABase has successfully served ByteDance's large-scale cloud environment, supporting a total workload that has achieved a peak QPS of over 13 billion and total storage exceeding 1 EB.
\end{abstract}




\maketitle

\section{Introduction}

Serverless NoSQL databases have emerged as pivotal technologies in cloud-native environments, supporting large-scale, highly available applications. These systems offer elastic and flexible data storage solutions without the need for infrastructure management, effectively meeting the demands of modern applications that require rapid deployment and significant elasticity. Cloud providers have utilized multi-tenant architectures in their serverless NoSQL databases. By co-locating different tenants within the same resource pool and sharing resources~\cite{jia2021systematic, bezemer2010multi}, this approach has shown substantial potential in maximizing elasticity and enhancing resource utilization. Several multi-tenant NoSQL databases have been deployed in production environments, including Amazon DynamoDB~\cite{elhemali_ATC_DynamoDB_2022}, Microsoft CosmosDB~\cite{azure_cosmos_db}, and Google Firestore~\cite{kesavan_ICDE_Firestore_2023}.

In large-scale cloud environments, a well-implemented multi-tenant database must address the challenges arising from diverse workloads and dynamically evolving data requests. For instance, ByteDance
operates across a broad array of business domains, such as e-commerce, search, social media, and AI services. 
Similar to other large internet corporations, each of ByteDance's domains demonstrates significant workload diversity, characterized by differing requirements for resources such as throughput, storage, and cache hit ratios across different business scenarios.
Additionally, workload dynamism is reflected in rapid changes in resource consumption by tenants, like throughput surges and sharp drops in cache hit ratios.
We will analyze these aspects in detail in Section~\ref{sec:bytedance_business}.

As a result, to be practical in such a large-scale cloud environment, a multi-tenant NoSQL serverless database must fulfill a diverse range of roles—serving as a high-speed cache, a large-capacity persistent storage, and a foundational layer for other systems—while also meeting immense performance requirements. For example, in ByteDance, we must manage a total peak QPS (queries per second) exceeding 13 billion and storage capacities surpassing 1 exabyte (EB). For individual tenants, the maximum QPS can reach 450 million, and the highest storage capacity exceeds 11 petabytes (PB).
After careful research, we concluded that existing multi-tenant NoSQL databases may not adequately meet our business needs due to their insufficient traffic capacities and designs intended for limited scenarios. 
We will elaborate on this in Section~\ref{sec:related_works}.
To effectively manage these highly \textbf{diverse} and \textbf{dynamic} workloads, we developed ABase, a multi-tenant NoSQL database system at ByteDance. 
In the course of designing, implementing, and maintaining ABase, we have identified three unique yet significant challenges that are typically encountered in multi-tenant NoSQL systems within large-scale cloud environments:

\begin{table*}[tbp]
    \centering
    \caption{Diverse application scenarios and workload characteristics of ABase in ByteDance business.}
    \label{tab:abase-business}
    \begin{tabularx}{\textwidth}{>{\hsize=1.2\hsize}X>{\hsize=1.1\hsize}X>{\hsize=0.7\hsize}X>{\hsize=0.7\hsize}X>{\hsize=0.6\hsize}X>{\hsize=0.6\hsize}X>{\hsize=0.6\hsize}X>{\hsize=0.6\hsize}X}
    \toprule
    \textbf{Business lines} & \textbf{Workloads} & \textbf{Normalized\newline Throughput} & \textbf{Normalized\newline Storage} & \textbf{Ratio of\newline Cache Hit} & \textbf{Ratio of \newline Read } & \textbf{Mean K-V\newline Data Size} & \textbf{Common\newline TTLs} \\
    \midrule
    Social Media (Douyin)    & Comment        & 250                   & 125                & 54\%               & 100\%         & 0.1KB                  & -  \\
    Social Media (Douyin)    & Direct message & 25                    & 678                & 74\%               & 100\%         & 1KB                   & -  \\
    E-Commerce               & Metadata tags  & 575                   & 42                 & 92\%               & 100\%         & 1KB                   & -  \\
    Search                   & Forward sorted data & 1500              & 63                 & 99\%               & 100\%         & 1KB                   & -  \\
    Advertisement            & For message joiner  & 2750              & 938                & 18\%               & 25\%          & 10KB                  & 3 hours \\
    Recommendation           & For deduplication   & 5325              & 625                & 76\%               & 50\%          & 2KB                   & 15 days \\
    Large Language Model    & Remote K-V Cache            & 10000 & 5760               & 0\%  & 85\%          & 5MB & 1 days \\
    \bottomrule
    \end{tabularx}
\end{table*}

\textbf{Challenge 1}:
In high-speed caching scenarios, tenants require frequent access to recently-updated data with low latency. ABase incorporates both proxy and data node caches to satisfy this need.
A request hitting the cache significantly alters both the execution process and resource consumption. 
However, exactly predicting whether a request will access the cache is challenging, introducing uncertainty and complexity into the performance isolation mechanism.
For example, requests that hit the proxy cache are directly returned without entering the data node, while those that hit the data node cache consume only CPU and memory resources, without disk I/O resources. This necessitates the systematic integration of caching considerations into the isolation mechanism\footnote{In this paper, "isolation" specifically refers to performance isolation between tenants, not to the isolation property in relational database ACID guarantees.}.
Previously multi-tenant NoSQL databases~\cite{elhemali_ATC_DynamoDB_2022, azure_cosmos_db, kesavan_ICDE_Firestore_2023} did not discuss cache impact on isolation or propose cache-aware isolation mechanisms.

\textbf{Challenge 2}:
Traffic patterns change dynamically, reflected in two aspects:
First, as tenant traffic trends increase, the pre-applied resources (termed ``quota'') may become exhausted, thereby triggering throttling. Conversely, a sustained traffic decrease in tenant traffic typically results in the wastage of these resource quotas.
Second, even when traffic volume remains constant, changes in access distribution can lead to hot key pressure if requests concentrate on a few keys, or significant drops in cache hit ratios if the accesses become dispersed.
To our knowledge, previous multi-tenant NoSQL systems have not integrated temporal forecasting as we have, leaving the hot key issue unresolved.

\textbf{Challenge 3}:
Each tenant has differing requirements on request traffic and storage; if the layout of tenant data is not carefully planned, it can lead to imbalanced resource utilization within and across data nodes, thereby limiting overall resource utilization. For instance, if all tenants assigned to a certain data node are storage-heavy but have low traffic, this can result in high disk resource utilization while CPU resources remain idle. Although this challenge is common in large-scale cloud environments, previously reported multi-tenant NoSQL serverless systems have not provided explicit implementations or algorithms to address it.

To address these challenges in large-scale cloud environments, we have made the following innovative contributions in ABase: 

\textbf{(1). Cache-Aware Isolation Mechanism (Challenge 1)}: 
We designed a cache-aware request unit (RU) that incorporates the cache hit ratio into RU computation, and introduced request restrictions at both the proxy and data node layers to control traffic.
Within the data node, we implemented a dual-layer Weighted Fair Queuing (WFQ).
The CPU-WFQ schedules requests and checks their existence in the data node cache; upon a cache miss, the I/O-WFQ further schedules requests to retrieve data from the disk layer.

\textbf{(2). Hierarchical Caching Mechanism (Challenge 2)}: 
At the data node layer, we implemented a cache based on size-aware LRU, employing individual eviction policies for items of different sizes to improve the cache hit ratio. At the proxy layer, we implemented a cache based on auto-updated LRU, along with a limited fan-out hash strategy, to effectively address both hot keys and sharp declines in cache hit rates.

\textbf{(3). Predictive Autoscaling Policy (Challenge 2)}: 
We outline the challenges of workload forecasting in ABase, such as non-periodic bursts, period diversity, and trend variability. We then propose an ensemble-based forecasting solution that combines the adaptive-periodic Prophet model with historical averages to achieve accurate predictions and elastic adjustments.

\textbf{(4). Multi-Resource Rescheduling Algorithm (Challenge 3)}: 
Considering the trade off between efficiency and effectiveness, we propose a heuristic multi-resource rescheduling algorithm to balance traffic and storage utilization across data nodes within a resource pool. We further extend this algorithm to support the data balancing across multiple resource pools.

\textbf{(5). Production Analysis, Evaluation and Lessons}:
We conduct comprehensive experiments to validate our contributions in large-scale cloud environments and provide detailed business analysis along with practical operational lessons. ABase has been examined for managing ten-billion-level QPS and exabyte-level data storage. We believe the insights offered in this paper will prove valuable to readers.

The rest of this paper is organized as follows: Section~\ref{sec:bytedance_business} introduces the business scenarios and workloads in a large-scale cloud environment, exemplified by ByteDance. Section~\ref{sec:architecture} provides an overview of ABase's architecture and design principles. Section~\ref{sec:implementation} details the system implementation of ABase. Section~\ref{sec:workload_management} discusses workload management strategies, including predictive autoscaling and rescheduling algorithms. Section~\ref{sec:experiments} presents experimental results validating the system's performance. Section~\ref{sec:lessons} discusses key lessons learned from ABase's lifecycle. Section~\ref{sec:related_works} reviews related work and our analysis. Finally, Section~\ref{sec:conclusion} concludes the paper.

\section{Background}
\label{sec:bytedance_business}

In this section, we sketch an overview of workload diversity and dynamism using ByteDance as a case study. We believe these phenomena are also applicable to other large-scale cloud environments.

\subsection{Diversity}

ABase supports a broad spectrum of business lines, and Table~\ref{tab:abase-business} reveals significant workload diversity within and across various business lines. 
For clarity, throughput and storage metrics have been normalized according to an empirical standard unit. If the normalized throughput and storage metrics are comparable, this indicates a balanced demand for CPU and disk resources in this workload.
The complexity of these diverse business requirements stems from variations in data characteristics and the ways in which ABase is utilized. 
First, considering the diversity within business lines, in the Social Media sector (Douyin), two workloads for comments and direct messages require different throughput-to-storage ratios (250:125 and 25:678, respectively). 
Next, considering the diversity across business lines, E-commerce and Search sectors demonstrate a preference for higher throughput over storage, with cache hit ratios exceeding 90\% due to frequent reads of hot data and few updates. The Advertisement and Recommendation sectors necessitate high throughput and storage capacities. Notably, the cache hit ratio for the Advertisement workload is a mere 18\%, which can be attributed to the specific application of the advertisement message joiner, where most data is read only once after being written. 
ABase supports large language models (LLM) by providing a remote caching store for kv-cache data, facilitating the caching of key-value results from token sequences to reduce costly recalculations during the generation of new tokens. These workloads demand throughput and storage capacities significantly higher than typical applications, normalized at 10000 and 5760, respectively.
LLM's cache ratio is 0, as it bypasses caching to directly process data from underlying logs, optimizing network bandwidth and query speed.

We further illustrate the read ratios (reflecting the operation distribution), K-V data sizes, and common TTLs (time-to-live) of these workloads. 
Most workloads in Table~\ref{tab:abase-business} are read-heavy or balanced, but ABase also serves write-heavy scenarios, such as the advertising business. K-V data sizes vary significantly across workloads. For instance, document and advertisement message data sizes are 7 KB and 10 KB respectively, while social media comments are only about 0.1 KB. 
Some businesses exhibit typical access patterns, with common TTLs set at about 3 hours for advertisements, 15 days for recommendations, and 1 day for language models.

\subsection{Dynamism}
Based on ABase experiences at ByteDance, we identified three challenging workload dynamism scenarios:

\textbf{(1). Throughput sharply increases}: During annual shopping events such as the Double-11 Shopping Festival and Black Friday, we observe rapid and significant increases in throughput among many tenants. These escalating workloads come from various sectors, including e-commerce, advertising, and search, with traffic peaks concentrated within a single week.

\textbf{(2). Cache hit ratio sharply declines}: A rapid increase in throughput can significantly reduce cache hit ratios. Additionally, even when throughput is stable, a business's cache hit ratio may still experience a substantial decline, often due to shifts in access patterns, such as ad hoc access to large volumes of older, cold data.

\textbf{(3). Emergence of Hot Keys}: In sectors like social media and search, hot events often cause a small amount of data to be heavily accessed. In multi-tenant architectures, the hot key issue is considered a "last mile" problem~\cite{elhemali_ATC_DynamoDB_2022} because the system must accommodate heavy traffic with a limited number of data nodes and cannot resolve this through data partitioning and migration.

Workload dynamism is also evident in other scenarios. Adjustments to the data TTL (time-to-live) can lead to rapid fluctuations in storage capacity. For tenants across multiple data centers, changes in traffic strategy can cause rapid shifts in traffic, read-write ratios, and cache hit ratios. Workload dynamism manifests in varying resource consumption among tenants, posing challenges to elasticity, load balancing, and tenant isolation.

\section{Architecture}
\label{sec:architecture}

\begin{figure*}[htbp]
    \centering
    \includegraphics[width=0.98\linewidth]{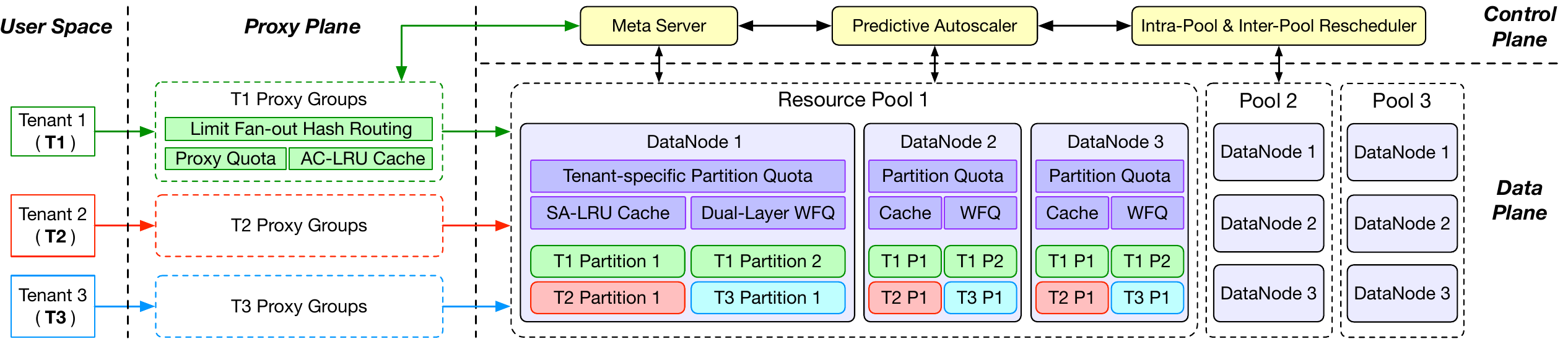}
    \caption{ABase multi-tenant architecture. }
    \label{fig:arch_overview} 
\end{figure*}

\subsection{Data Model and Design Rationale}

ABase supports the Redis protocol to ease adoption for users familiar with Redis, and enables eventual consistency.
As shown in Figure~\ref{fig:arch_overview}, the ABase system comprises a series of \textit{resource pools}, each managing a suite of \textit{tenants}. A tenant can create several key-value tables, where each \textit{table} is composed of numerous items, each identified by a unique key. The data belonging to a tenant are uniformly allocated into several contiguous and disjoint \textit{partitions} accordingly. Each partition generates multiple \textit{replicas} across various Availability Zones (AZs), thus enhancing availability and security. 



ABase introduces a resource pooling concept that distributes multiple tenants' data across individual physical machines, forming a vast resource pool. 
Data partitioning plays a key role in the multi-tenant architecture. ABase divides each tenant's data into multiple non-overlapping partitions and strives to distribute these partitions across different machines within the same resource pool. 
The multi-tenant architecture can utilize workload diversity, allowing tenants with different resource demands to be co-located, thereby enhancing machine resource utilization. 


ABase isolation adheres to two fundamental principles: Firstly, the design of isolation must consider the impact of caching, encompassing RU, quota, and the request queue. Secondly, traffic control for individual tenants should be prioritized before the traffic reaches the shared request queue. Using dedicated resources, individual restrictions are designed to block excessive traffic at the single-tenant stage, thereby enabling the shared request queue to focus on fair and efficient request processing among multiple tenants under moderate traffic pressure, rather than rejecting enormous requests.

Continuous growth in traffic may trigger tenant throttling, while dynamic changes in tenant traffic and storage can unbalance the load across ABase's data nodes. ABase adopts predictive scaling to maintain a modest ratio between applied resources and actual utilized resources, and deploys rescheduling algorithms to periodically balance tenant replicas both within and across resource pools. However, neither of these measures addresses the issue of hot keys, which we address by introducing an innovative caching strategy.
    
\subsection{Multi-Tenant Architecture}

\noindent\textbf{Architecture Overview}: Figure~\ref{fig:arch_overview} depicts the overall architecture of ABase. ABase comprises three parts: the control plane is a centralized management component that administers a series of resource pools for traffic management, scaling, and rescheduling. The data plane comprises several resource pools. Within each pool, numerous DataNodes manage multiple partitions for different tenants. The proxy plane contains tenant proxies, responsible for routing tenant requests to the relevant data nodes.

\textbf{Control Plane} comprises meta server, autoscaler, and rescheduler. The meta server serves as the centralized management module for ABase, tasked with managing global metadata, monitoring resource pool health, repairing data nodes, and overseeing the scaling and migration of data partitions. 
The autoscaler collects metrics on tenant RU and storage utilization, making tenant scaling decisions based on time-series forecasting. 
The rescheduler uses the same metrics to trigger rescheduling events, migrating replicas both within and between resource pools.

\textbf{Data Plane} contains multiple resource pools, each comprising multiple DataNodes. Each DataNode is allocated a physical disk along with corresponding CPU resources and manages multiple partition replicas for a diverse range of tenants. 
DataNodes handle partition-layer traffic control based on each tenant's specific partition quota. DataNodes are equipped with a cache that utilizes Size-Aware LRU (SA-LRU) and a fine-grained Weighted Fair Queueing (WFQ) module, together ensuring Quality of Service (QoS) in multi-tenant environments.

\textbf{Proxy Plane} consists of proxies belonging to various tenants. The primary function of the proxy is to route requests. Upon receiving a client-initiated request, the proxy communicates with the MetaServer to obtain essential routing details for tenant partitions to facilitate subsequent request retransmission. Proxies conduct the proxy-layer traffic control based on each tenant's specific proxy quota. To further enhance ABase's ability to defend against cache hit dynamism and hot key issues, proxies are equipped with a cache based on active-update LRU (AC-LRU), and proxies for each tenant are organized into proxy groups that adopt a Limit Fan-out Hash Routing strategy to enhance the cache hit ratio.

\subsection{Recovery and Robustness}
\label{sub:recovery}

The multi-tenant architecture of ABase exhibits superior recovery capabilities over single-tenant designs, facilitated by parallel processing and resource pooling. When a DataNode fails, the MetaServer coordinates parallel replica reconstruction across operational nodes, thereby effectively utilizing multi-node disk I/O bandwidth to accelerate recovery. This distributed approach eliminates a fundamental constraint in single-tenant systems, wherein the complete replica restoration on a single replacement node is significantly constrained by its disk I/O limitations. Moreover, the architecture maintains robustness while achieving higher resource utilization through its shared resource pool. For example, in single-tenant system with 3 replicas, resource utilization must remain below 2/3 to accommodate potential 3/2 workload spikes during single node failure. The multi-tenant design mitigates this impact through N-node redundancy, where load redistribution results in only a 1/N utilization increase on surviving nodes, thus enabling sustainable high utilization without compromising fault tolerance.

\section{System Implementation}
\label{sec:implementation}

\subsection{Normalized Request Unit}
\label{sub:ru}

Request Units (RUs) are widely employed in serverless databases to direct user focus towards request throughput demands~\cite{elhemali_ATC_DynamoDB_2022,azure_cosmos_db} and to abstract from underlying hardware complexities. In ABase, RUs are not only crucial for billing but also constitute a key component of the isolation mechanism by quantifying a request's consumption of CPU, memory, and disk I/O. 
We demonstrate how ABase tailors RU estimation to different request types, ensuring that RUs closely reflect the actual resource consumption of operations, while taking into account the impact of caching on resource consumption.

\textbf{Write Operations:} For write operations, the value size of the written item $S_{\text{write}}$ is typically known, which facilitates a straightforward computation of $RU_{\text{write}} = S_{\text{write}} / U$, where $U$ is the unit byte size, empirically set to 2KB. Importantly, considering ABase's replication mechanism, a single user write request translates into one direct write operation and $r-1$ synchronization operations to other replicas (where $r$ is the number of replicas), resulting in a total charge of $r \cdot RU_{\text{write}}$.

\textbf{Read Operations:} Since the value size and cache hit status of read operations are not predetermined, we estimate the size of upcoming reads, $\mathbb{E}[S_{\text{read}}]$, and cache hit ratios, $\mathbb{E}[R_{\text{hit}}]$, using a moving average of the last $k$ requests. 
We employ $\mathbb{E}[S_{\text{read}}]$ for traffic control, detailed in Section~\ref{sub:quota}, and charge based on the actual size returned. Requests that hit the proxy cache are directly returned without throttling or charges. In summary, the formula for estimating read costs is 
$RU_{\text{read}} = \mathbb{E}[S_{\text{read}}] \times (1 - \mathbb{E}[R_{\text{hit}}]) / U$.

\textbf{Complex Read Operations:} Challenges in estimating RUs for complex operations stem from the unpredictable number of items a request may scan (e.g., \text{HLen} (the number of fields in a hash table)) and the intricate multi-stage procedures involved in requests (e.g., \text{HGetAll} (a command to retrieve all fields and values in a hash table)). 
To estimate \text{HLen}, we use historical data on the length of the HashSet, and for \text{HGetAll}, we decompose the operation into \text{HLen} followed by a scan, calculating the RU for each stage separately.

\subsection{Hierarchical Request Restriction}
\label{sub:quota}

ABase implements a hierarchical request restriction strategy, divided into \textbf{proxy-level} and \textbf{partition-level}. As shown in Figure~\ref{fig:node_qos_wfq}, each tenant is assigned a dedicated set of proxies, proportional to its allocated quota. Proxies forward their respective requests to the DataNodes. 
Note that, requests that hit the proxy cache do not consume any proxy quota.
DataNodes route the requests to a request queue, filtering out those that exceed predefined quotas.
Remaining valid requests are processed by the subsequent Dual-Layer WFQ module, which will be discussed in Section~\ref{sub:wfq}.

At the \textbf{proxy level}, the primary duty of the proxy is to prevent the total RUs from surpassing the tenant quota. Unlike DynamoDB, which requires real-time interactions between request routers and the Global Admission Control, ABase Proxy employs an asynchronous traffic control strategy to minimize dependencies between proxies and the centralized MetaServer. Each proxy receives a specific \textit{proxy\_quota}, calculated by dividing the tenant quota by the number of proxies, allowing them to process up to double this quota autonomously. To maintain the tenant's total traffic across all proxies within the set tenant quota, the MetaServer continuously monitors each proxy's traffic and, if exceeded, directs the proxies to revert to their standard \textit{proxy\_quota}.

Another responsibility of the proxy is to shield tenants on DataNodes — which are shared among multiple tenants — from the impact of co-tenants' burst traffic. When the traffic of certain tenants significantly escalates, the proxy designated for this tenant can reject excess traffic, thereby preventing requests from reaching the DataNodes. This avoidance helps reduce the extensive resource consumption that would occur if DataNodes were to handle and reject these requests, thus safeguarding the stability of other tenants.

At the \textbf{partition level}, DataNodes reject requests that exceed the maximum allowed quota of a partition at the entry point, namely the request queue. In DynamoDB, a partition is allowed to consume the entire tenant quota. However, under extreme conditions, this flexibility inevitably leads to mutual interference among co-tenants. 
Elevated loads on specific tenant partitions may deplete resources, potentially leading to service degradation as traffic surges in previously low-load tenants. 
ABase explicitly introduces a \textit{partition\_quota}, defined as the tenant quota divided by the number of partitions, ensuring that no single partition surpasses \textbf{three times} its \textit{partition\_quota}. This restriction is reasonable because ABase organizes all items in an hash partition, thus ensuring that each partition is likely to experience even traffic. We will discuss hot key optimization in Section~\ref{sub:cache}.

\subsection{Dual-Layer Weighted Fair Queueing}
\label{sub:wfq}
 
In the ABase system, each ABase DataNode may host partitions belonging to various tenants. To ensure fair and efficient handling of requests from various tenants, we designed a fine-grained, dual-layer Weighted Fair Queueing (WFQ) mechanism.

\begin{figure}[tbp]
    \centering
    \includegraphics[width=1\linewidth]{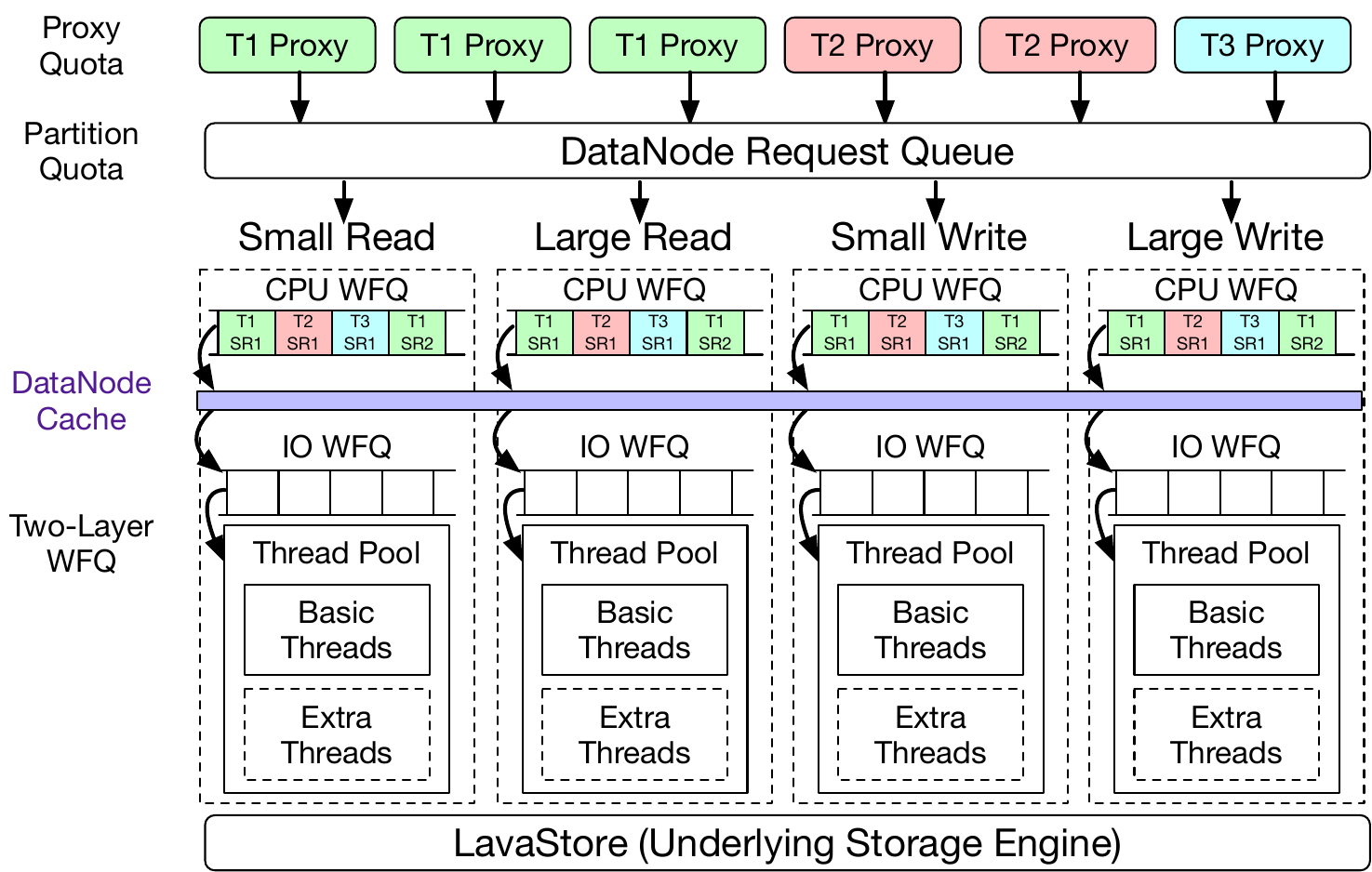}
    \caption{Cache-aware performance isolation.
    }
    \label{fig:node_qos_wfq}
\end{figure}

As noted in 2DFQ~\cite{mace_SIGCOMM16_2DFQ_2016}, a fair and efficient WFQ is expected to prevent interference between heavyweight and lightweight requests. To achieve this objective, we have implemented a straightforward yet robust approach. As illustrated in Figure~\ref{fig:node_qos_wfq}, all requests are categorized into four independent dual-layer WFQs based on their type (read/write) and their size (large/small). This categorization has proven effective in practice, as it ensures closely matched request latencies within each queue type.

Resource consumption by requests depends on whether they hit the DataNode cache. 
Referencing SQLVM~\cite{narasayya_CIDR_SQLVM_cidr13_2013,das_VLDB_SQLVM_vldb13_2013}, we designed dual-layer queues, including CPU-WFQ and I/O-WFQ. Requests first enter the upper CPU-WFQ for processing. If a request hits the cache, it can be directly returned; otherwise, it proceeds to the lower I/O-WFQ. The I/O-WFQ uses a group of basic threads to handle normal requests, and employs additional threads to handle requests from other tenants when basic threads are fully occupied by a single tenant.

WFQ acts as a min-heap to prioritize requests with the customized smallest virtual finish time (VFT). ABase has elaborately designed the VFT to ensure efficient and fair execution among various tenant requests. The VFT of a request is formulated as follows:
\begin{align*}
\text{wReqCost}(Q_i) &= \frac{\text{Cost}(Q_i)}{\text{wPartition}(Q_i)} = \frac{\text{Cost}(Q_i)}{Q_i/\sum {Q_p}}\\
\text{VFT}(Q_i) &= \text{preVFT}_{T_i} + \text{wReqCost}(Q_i)
\end{align*}
A request's cost is weighted according to its partition quota, where a higher proportion of the partition quota in the DataNode (denoted by $\text{wPartition}$) leads to a higher weight cost (denoted by $\text{wReqCost}$). Moreover, the VFT for all requests from the same tenant is cumulative, thereby preventing scenarios where a single tenant's requests are consistently prioritized high, even if that tenant has a larger partition quota or lower request costs.

Furthermore, to address challenges encountered in practical deployments, we introduced the following enhancements:

\textbf{Rule 1}: We defined different $\textit{Cost}(Q_i)$ for requests in CPU-WFQ and I/O-WFQ. For CPU-WFQ, costs are based on RU, while for I/O-WFQ, they are determined by the request's IOPS. This is based on observing that in ABase, a single I/O operation generally has a similar execution time, regardless of request detail.

\textbf{Rule 2}: In CPU-WFQ, concurrency limits are enforced on both read and write requests to ensure stable latency~\cite{apache_brpc}. For write operations, in addition to managing concurrency, we impose a ceiling on the total RU to enhance stability of write latencies during compaction and garbage collection processes in the LavaStore storage engine~\cite{wang2024lavastore}. This strategy is essential in preventing significant throughput oscillations within the storage engine.

\textbf{Rule 3}: All requests from a single tenant can occupy at most 90\% of the CPU-WFQ resources, even if that tenant has a substantial partition quota. This rule is designed to prevent significant delays in other tenants' requests during traffic bursts from a single tenant.

\textbf{Rule 4}: If all \textit{basic threads} in the I/O-WFQ thread pool are monopolized by tasks from one tenant, we temporarily increase \textit{extra threads} to handle tasks from other tenants. This strategy relies on the assumption that simultaneous high traffic from two tenants on a DataNode is unlikely; therefore, a few additional threads are sufficient to manage such conflicts.


\subsection{Dual-Layer Caching}
\label{sub:cache}

\noindent{\textbf{Proxy-Layer Cache}}:
Hot key management presents a critical challenge for key-value databases, particularly during high-traffic scenarios such as major promotional events. Previously discussed techniques, such as partition splitting and rescheduling, have proven inadequate to manage the strain placed by high-frequency access to a few hot keys on a single data node~\cite{elhemali_ATC_DynamoDB_2022}. Traditional caching approaches face limitations due to proxy memory constraints, typically less than 10GB, leading to frequent cache evictions and suboptimal hit ratios under random routing schemes.

To address these challenges, we propose a dual-component optimization framework comprising a proxy-side cache module and client routing strategy. The proxy layer implements an Active-Update LRU (\textbf{AU-LRU}) mechanism that bypasses DataNode accesses if the proxy cache is hit. An active-update mechanism is applied to address potential spikes in requests due to expired cache entries. It automatically refreshes hot keys as they near expiration, thus maintaining the timeliness and continuity of the cached data.
On the client side, we adopt a \textbf{limited fan-out hash strategy} to determine the destination of requests. The tenant's $N$ proxies are divided into $n$ groups. When a tenant accumulates a list of requests, each request is hashed to one of the $n$ ProxyGroups using a custom hashing function. This group of requests will randomly choose one proxy from this hashed ProxyGroup to send out.
By carefully adjusting $n$, tenants can optimize the balance between hit ratio and hot key pressure. Because each proxy receives $1/n$ of the total requests, a larger $n$ results in a higher cache hit ratio for each proxy. During hot key events, selecting a smaller $n$ value facilitates load distribution across a larger number of proxies ($=N/n$).

\noindent{\textbf{DataNode-Layer Cache}}: Workload diversity necessitates the management of a broad spectrum of key-value data sizes. Inspired by these considerations, we developed a DataNode-layer cache that utilizes the enhanced Size-Aware LRU strategy (SA-LRU). This strategy, SA-LRU, strategically evicts data that occupies more memory while yielding fewer cache hits, effectively managing memory more efficiently. By prioritizing the retention of smaller-sized data, which typically incurs lower access costs, SA-LRU not only optimizes resource utilization but also enhances the overall cache hit ratio.

\section{Workload Management}
\label{sec:workload_management}

\subsection{Predictive AutoScaling}
\label{sub:scaling_policy}

Algorithm~\ref{alg:scaling} shows the details of the ABase scaling policy. Details on workload forecasting and resource rescheduling are further explored in subsequent sections (Section~\ref{sub:forecasting} and Section~\ref{sub:rescheduling}).
In the ABase system, quotas are categorized into RU (Request Unit) and Storage, each allowing for independent scaling by tenants. Suppose a tenant with tenant quota $Q_T$, number of partitions $N$, and its partition quota $Q_P$. We forecast the resource usage in next 7 days based on a 30-day historical series as $U_{max}$. When the forecasted usage exceeds the upper threshold (0.85) or falls below the lower threshold (0.65) of the tenant quota, scaling up or down is triggered accordingly. After scaling up, if the partition quota exceeds the quota upper bound $\text{UP}$, a partition split is triggered; after scaling down, we ensure that the partition quota does not fall below the quota lower bound $\text{LOWER}$ to accommodate occasional traffic bursts from tenants. Scaling operations result in changes to the distribution of partition quotas and usage; thus, ABase continuously invokes the rescheduling strategy to balance the utilization of DataNodes within the resource pool.

\begin{algorithm}
\caption{ABase Scaling Policy}
\label{alg:scaling}
\begin{algorithmic}[1]
\Require $Q_T$, $N$, $U_{max}$

\If{$U_{max} > 0.85 \times Q_T$}
    \State $Q_T \leftarrow U_{max} / 0.65$; 
    \State $Q_P \leftarrow Q_T / N$;
    \If{$Q_P > \text{UP}$}
        \State Trigger partition split so that $Q_P \leftarrow 0.5 \times Q_P$
    \EndIf

\ElsIf{$U_{max} < 0.65 \times Q_T$ and not scaled in last 7 days}
    \State $Q_T \leftarrow U_{max} / 0.65$; 
    \State $Q_P \leftarrow \max(Q_T / N, \text{LOWER})$

\EndIf

\State Invoking rescheduling strategy periodically.
\State Continue to forecast the max usage $U_{max}$ in next 7 days.

\end{algorithmic}
\end{algorithm}

\subsection{Workload Forecasting}
\label{sub:forecasting}

The workload forecasting module is crucial in the ABase autoscaling. It processes resource usage metrics from the past 30 days, downsampled to 1-hour intervals, along with their quota records, and predicts the resource usage trends for the next 7 days to inform scaling decisions.
Although time-series based scaling strategies are commonly utilized in cloud services~\cite{poppe_VLDB_Moneyball_2022,poppe_VLDB_Seagull_2020,zhou_AAAI_AHPA_2023,qian_ICDE_RobustScaler_2022,pan_VLDB_MagicScaler_2023,rzadca_EuroSys_Autopilot_2020,curino2011workload,novakovic2013deepdive,xue_KDD22_MetaReinforcementLearning_2022}, ABase faces several complex challenges in practice:

\textbf{Issue 1: Sporadic Bursts and Metric Noise}: ABase must be cautious with scaling operations, which involve costly processes such as partition migration and resource pool scaling. Sporadic bursts, which may be ad-hoc and temporary, should not trigger unnecessary upscaling. Furthermore, metrics erroneously recorded during partition migrations or master node transitions can lead to their misinterpretation as transient bursts.

\textbf{Issue 2: Period Diversity and Trend Variability}: 
The periodicity of the ABase workload is highly diverse. Apart from standard daily and weekly cycles, it includes various uncommon periods, such as 3.5 days, often attributed to specific tenant TTL configurations. Significant trend variations frequently occur within individual series, typically due to business adjustments and data cleaning.

\textbf{Issue 3: Consistent Non-periodic Bursts}: For some tenants, peaks occur daily at varying times without regular periodicity. These should not be dismissed as mere outliers. Accurately predicting these bursts' maximum value is crucial for appropriate ABase scaling decisions.

To address these issues, we have developed an \textit{ensemble-based forecasting solution}. In the preprocessing phase, we apply multi-metric collaboration for denoising. If Usage and Quota metrics simultaneously show spikes, these are considered noise and filtered out, as such simultaneous occurrences are nearly impossible in practice. Additionally, we use heuristic methods to eliminate sporadic peaks, likely due to accidental events, such as those appearing only once in the past 10 days. We also utilize change point detection methods to identify trend shifts, thereby focusing the forecasting algorithms more on recent data changes (\textbf{for Issue 1}).

During the forecasting phase, we initially use power spectral density (PSD)~\cite{tobar_NIPS_PSD_Period_2015} analysis to determine the time series' periodicity. Subsequently, we employ a weighted ensemble of predictions derived from both the Prophet~\cite{taylor__Prophet_2018} and historical average methods~\cite{sun_ICDE_SUFS_2023}. The Prophet model is effective for time series with clear trends and periods, while the historical average provides stable forecasts, especially suitable when trend changes are minimal (\textbf{for Issue 2}). For consistent non-periodic bursts, if the forecasts are significantly lower than historical input data, we directly use the most recent period's historical data for predictions to avoid unnecessary downscaling (\textbf{for Issue 3}).

We have alao investigated other deep learning-based methods, like TFT~\cite{lim_In_J_Forecast_TFT_2021}, AutoFormer~\cite{wu_NIPS_Autoformer_2021}, N-Beats~\cite{oreshkin_AppliedEnergy_NBEATS_2021} and N-Hits~\cite{challu_AAAI_NHITS_2023}. Although these models yield high-quality forecasts after pre-training, our ensemble-based approach maintains comparable precision and robustness, seamlessly adapting to new tenants with emerging trend characteristics without the need for retraining. 

\subsection{Workload Rescheduling}
\label{sub:rescheduling}
To address imbalanced DataNode utilization from diverse workloads, ABase incorporates a novel resource rescheduling module. This module uses a heuristic approach to balance efficiency and effectiveness, with two components: \textit{intra-pool}, focusing on reallocations within a single pool, and \textit{inter-pool}, managing reallocations across different pools to optimize resource utilization.

The \textbf{intra-pool rescheduling algorithm} primarily consists of two phases. The first phase aims to balance the replica distribution for each tenant, distributing the count of a tenant's replicas across DataNodes as evenly as possible, thus enhancing elasticity and robustness against failures. The second phase aims to balance resource utilization across all DataNodes within a resource pool, involving two resource dimensions (RU and storage) without compromising the previously established replica balance. Both phases use similar heuristic algorithms; for brevity, we next focus on a detailed explanation of the second phase, resource utilization rescheduling.

\textbf{(1). Load Indicator}: We characterize the resource load (e.g., RU, Storage) of a Replica (\textbf{RE}), DataNode (\textbf{DN}), Resource Pool (\textbf{RP}) as follows. First, the load of each replica is aggregated based on the hourly average, retaining load data from the past seven days. This data is then aggregated by taking the maximum value within the hour-of-day dimension to derive the load vector $RE^{ld} =(RE^{ld}_1,\cdots,RE^{ld}_{24})$. Note that the RU load incorporates the weighted factors of read RU, write RU and the cache hit ratio.

Second, the load vectors of all replicas on the DataNode or Resource Pool are summed and the maximum value of the resulting vector is computed. The specific calculation formula is as follows:
\begin{align*}
DN^{ld}(RP^{ld}) = \max_i \left( \sum_{RE \in DN(RP)} RE^{ld}_i \right) \quad \text{for } i \in \{1, 2, ..., 24\}
\end{align*}
where $DN^{ld}(RP^{ld})$ represents the resource load of the DataNode(Resource Pool). 

\textbf{(2). Optimal Load}: Considering the necessity to balance resource load across multiple dimensions (RU, Storage), the optimal load vector $(R,S)$ within a single resource pool is defined as follows:
\begin{align*}
\langle R, S \rangle = \left(\frac{RP_{ru}^{ld}}{RP_{ru}^{cap}}, \frac{RP_{sto}^{ld}}{RP_{sto}^{cap}}\right)
\end{align*}
where $RP_{ru}^{ld}(RP_{sto}^{ld})$ represents the RU (Storage) load of the resource pool, and $RP_{ru}^{cap}(RP_{sto}^{cap})$ represents the total RU (Storage) capacity of the resource pool.

\textbf{(3). Migration Gain}: To quantify the benefits of migrating a replica $RE$ to $Des\_DN$ (the Destination DataNode), we initially define the deviation between a DataNode's load with optimal load. We employ the L2-Norm Loss to evaluate this deviation as follows:
\begin{align*}
\mathcal{L}(DN)=\sqrt{(\frac{DN_{ru}^{ld}}{DN_{ru}^{cap}} -R)^2+(\frac{DN_{sto}^{ld}}{DN_{sto}^{cap}} -S)^2}
\end{align*}
where $DN_{ru}^{ld}(DN_{sto}^{ld})$ represents the RU (Storage) load of the DataNode, $DN_{ru}^{cap}(DN_{sto}^{cap})$ represents the total RU (Storage) capacity of the DataNode.

Therefore, when migrating the replica $RE$ from its current DataNode ($RE.DN$) to the destination DataNode ($Des\_DN$), we quantify the migration's gain by the reduction in maximum load across both nodes post-migration. A decrease in maximum load indicates a positive gain, signifying improved load distribution:
\begin{align*}
\mathcal{G}(RE, Des\_DN)=\max[\mathcal{L}(RE.DN),\mathcal{L}(Des\_DN)]-\\
\max[\mathcal{L}(RE.DN.Remove(RE)),\mathcal{L}(Des\_DN.Add(RE)]
\end{align*}
where $RE.DN.Remove(RE)$ represents $RE.DN$ removing $RE$, and $Des\_DN.Add(RE)$ represents $Des\_DN$ adding $RE$.

\textbf{(4). DataNode Division}: Workload rescheduling is based on the heuristic of migrating replicas from high-loaded DataNodes to low-loaded DataNodes. Specifically, DataNodes are divided into three groups based on their load levels: $\mathcal{S}_{L}$ (Low Load DataNodes), $\mathcal{S}_{M}$ (Medium Load DataNodes), $\mathcal{S}_{H}$ (High Load DataNodes). Using the RU load as a case study, the DataNodes are divided as follows:
\[
\left\{
\begin{aligned}
&\text{DN} \in \mathcal{S}_{L}, && \text{if} \ \ \frac{DN_{ru}^{ld}}{DN_{ru}^{cap}} \leq R-\theta \\
&\text{DN} \in \mathcal{S}_{M}, && \text{if} \ \ R-\theta < \frac{DN_{ru}^{ld}}{DN_{ru}^{cap}} \leq R \\
&\text{DN} \in \mathcal{S}_{H}, && \text{others}
\end{aligned}
\right.
\]
where $\theta$ is the manually set threshold, such as 5\%.

Algorithm~\ref{alg:rescheduling} outlines the intra-pool workload rescheduling process. For each resource type, we categorize DataNodes into three groups: $\mathcal{S}_{L}$, $\mathcal{S}_{M}$, and $\mathcal{S}_{H}$. The algorithm iterates over each high-load DataNode ($Src\_DN$) in $\mathcal{S}_{H}$, excluding those with ongoing replica migrations.
For each eligible $Src\_DN$, the algorithm examines each replica $RE$ on it. It then considers all low-load DataNodes ($DN$) in $\mathcal{S}_{L}$ that meet two criteria: $DN.CanPlace(RE)$, which preserves the uniform distribution of table replicas without overloading $DN$ into the high-load set $\mathcal{S}_{H}$, and $DN.IsMigrating(RE)$, which verifies the absence of ongoing replica migration on $DN$.
The process concludes by selecting the replica $RE_{move}$ and destination DataNode $Des\_Node$ that maximize the gain function $\mathcal{G}(RE, DN)$. A positive gain triggers the execution of the migration.

\begin{algorithm}[tbp]
    \caption{Intra-Pool Workload Rescheduling}
    \label{alg:rescheduling}
    \begin{algorithmic}[1]
    \Require $\{DataNodes\}$
    \For {resource in [RU, Storage]}:
    \State $\mathcal{S}_{L}$, $\mathcal{S}_{M}$, $\mathcal{S}_{H}$ = \textbf{Division}($\{DataNodes\}$, resource)
    \For {$Src\_DN$ in $\mathcal{S}_{H}$}:
      \If {$Src\_DN.IsMigrating == TRUE$}:
      \State \textbf{Continue}
      \EndIf
      \State $Gain_{best} = 0$
        \For {$RE$ in  $Src\_DN.replicas$}:
        
        \For {$DN$ in $\mathcal{S}_{L}$}:
        \If {$DN.CanPlace(RE) == FALSE$}:
        \State \textbf{Continue}
        \EndIf
        \If {$DN.IsMigrating == TRUE$}:
        \State \textbf{Continue}
        \EndIf
        
        \If {$\mathcal{G}(RE, DN) > Gain$}:
            \State {$Des\_DN = DN$}
            \State {$RE_{move} =  RE$}
            \State {$Gain_{best} = \mathcal{G}(RE, DN)$}
        \EndIf
        \EndFor
        \EndFor
        \If {$Gain_{best} > 0$}:
        \State \textbf{Migration}($RE_{move}$,  $Src\_DN$, $Des\_DN$)
        \State $Src\_DN.IsMigrating = TRUE$
        \State $Des\_DN.IsMigrating = TRUE$
        \EndIf
    \EndFor
    \EndFor
    \end{algorithmic}
\end{algorithm}

In terms of the \textbf{inter-pool rescheduling algorithm}, it primarily focuses on reallocating DataNodes between resource pools, which can be readily extended from the intra-pool algorithm. For example, to balance the resource utilization between two resource pools, $Pool_{H}$ (with higher load) and $Pool_{L}$ (with lower load), we tend to vacate a portion of the DataNodes from $Pool_{L}$ and reallocate them to $Pool_{H}$. Initially, we select some low-utilization DataNodes from $Pool_{L}$ and migrate replicas from these selected DataNodes to others within the same pool ($Pool_{L}$). Then, we reassign these vacated DataNodes to $Pool_{H}$. Finally, we invoke the intra-pool algorithm to re-balance the load within the two resource pools. 

\section{Experiments}
\label{sec:experiments}

\subsection{Production Statistics}

\begin{figure}[htbp]
    \centering
    \includegraphics[width=0.75\linewidth]{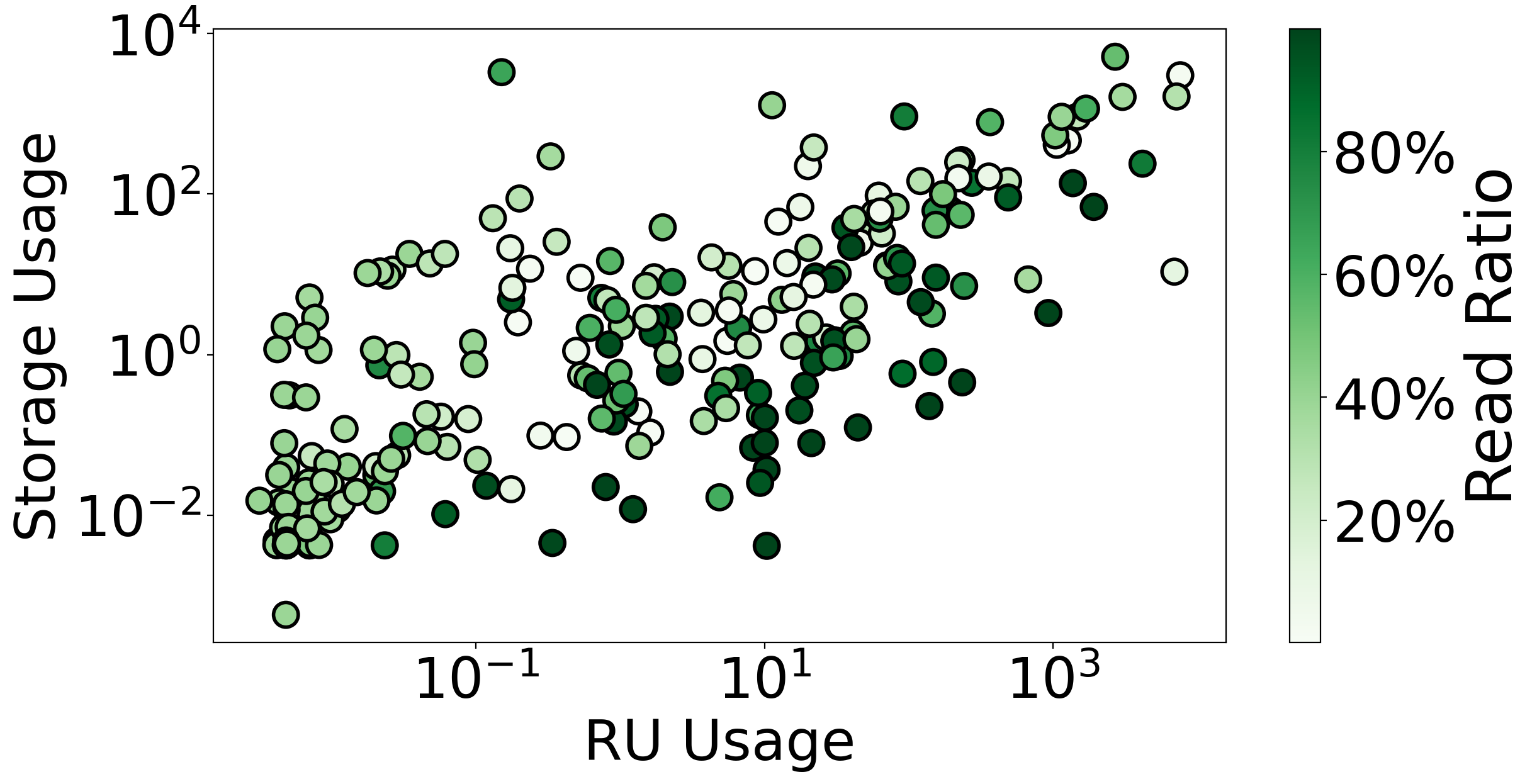}
    \caption{Distribution of tenants by RU, storage, and read ratio. Each point represents one tenant, normalized by median.
    }
    \label{fig:exp_diversity_rcu_storage}
    
\end{figure}

\noindent{\textbf{Diversity Analysis: }}
We present real production statistics from a specific resource pool at ByteDance in Figure~\ref{fig:exp_diversity_rcu_storage}, each circle represents a tenant in this pool, with the horizontal and vertical axes showing the tenant's average RU and storage usage over the past month, respectively. The color of each circle indicates the read operation ratio of the tenant, with darker colors indicating a higher read ratio. 
Generally, tenants with higher RU tend to have larger storage capacities, yet there are numerous cases exhibiting diverse RU/storage characteristics. 
In terms of the read ratio, it can be observed that tenants with a larger ratio of RU to storage (the lower right corner of Figure~\ref{fig:exp_diversity_rcu_storage}) tends to indicate a read-heavy workload.

\begin{figure}[tbp]
    \centering
    \begin{subfigure}{0.23\textwidth}
        \includegraphics[width=\textwidth]{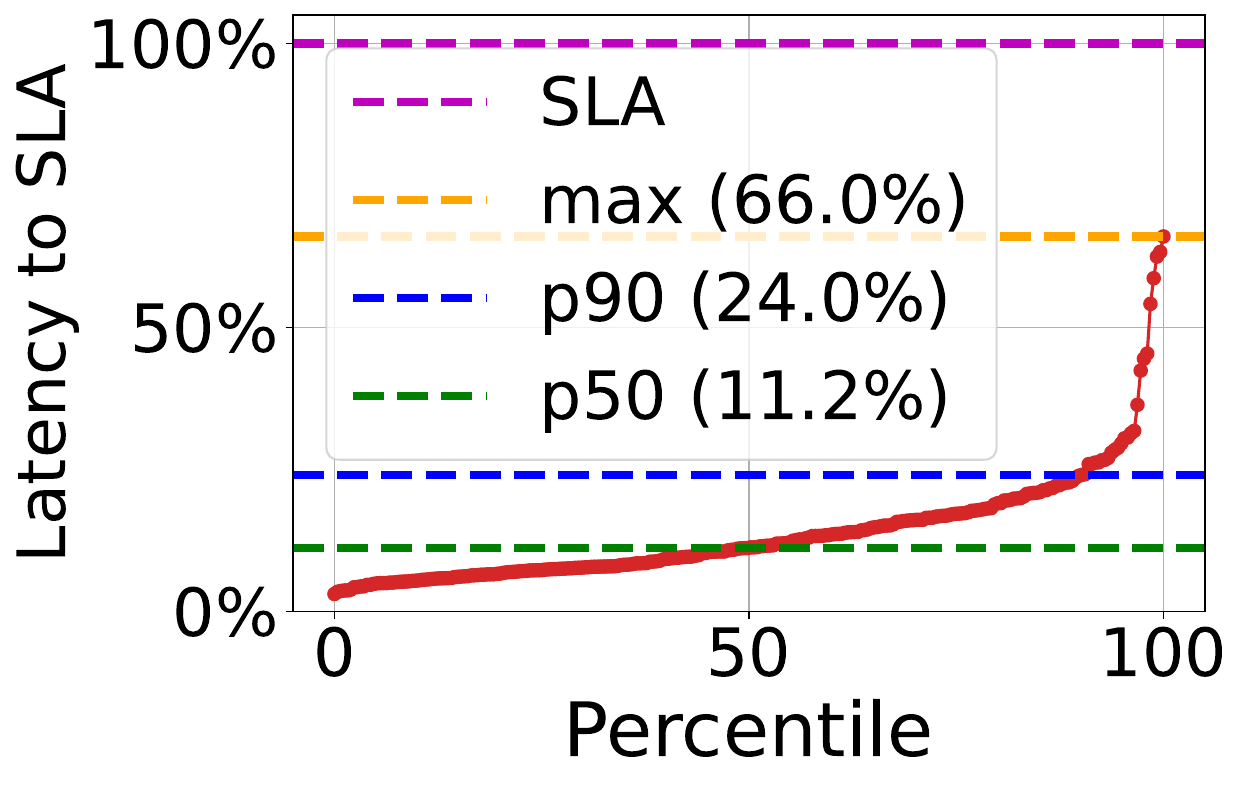}
        \caption{Latency to SLA}
        \label{subfig:exp_latency_percentile}
    \end{subfigure}
    \hfill
    \begin{subfigure}{0.23\textwidth}
        \includegraphics[width=\textwidth]{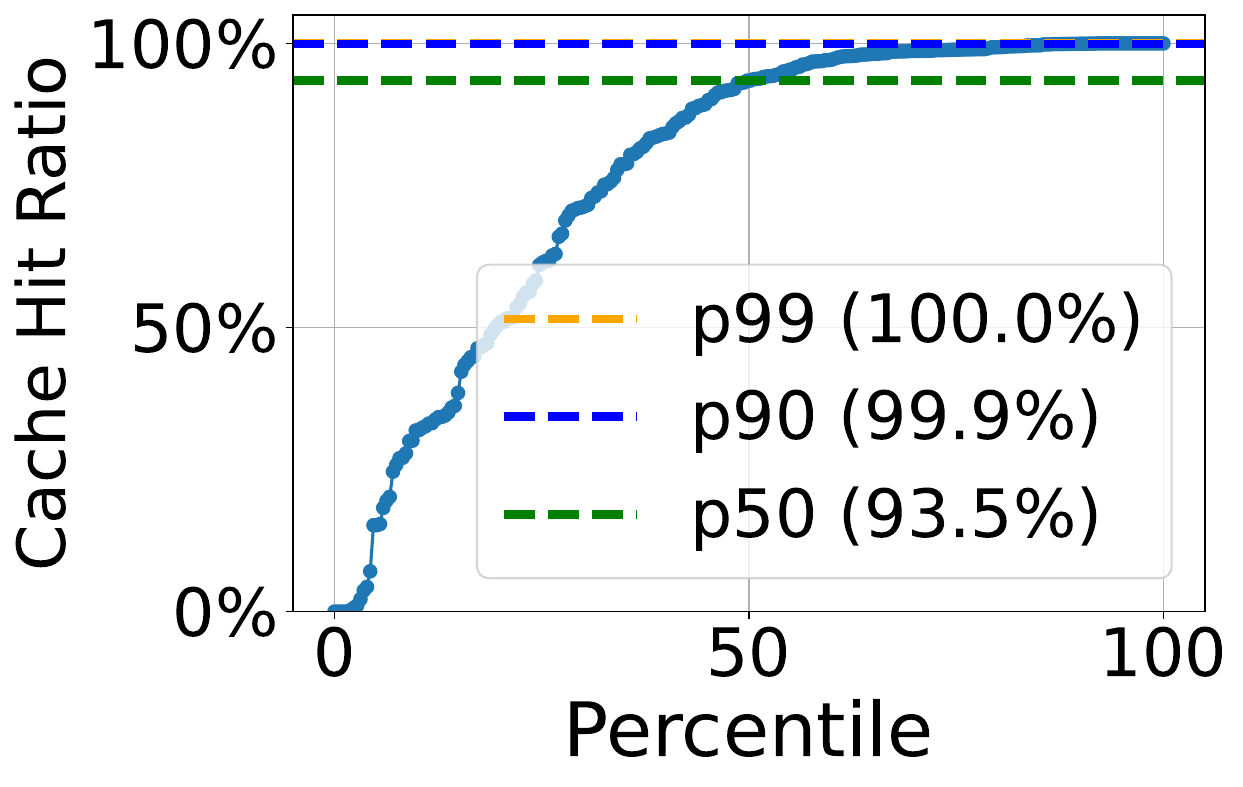}
        \caption{Cache Hit Ratio}
        \label{subfig:exp_cache_hit_ratio_perentile}
    \end{subfigure}
    \begin{subfigure}{0.23\textwidth}
        \includegraphics[width=\textwidth]{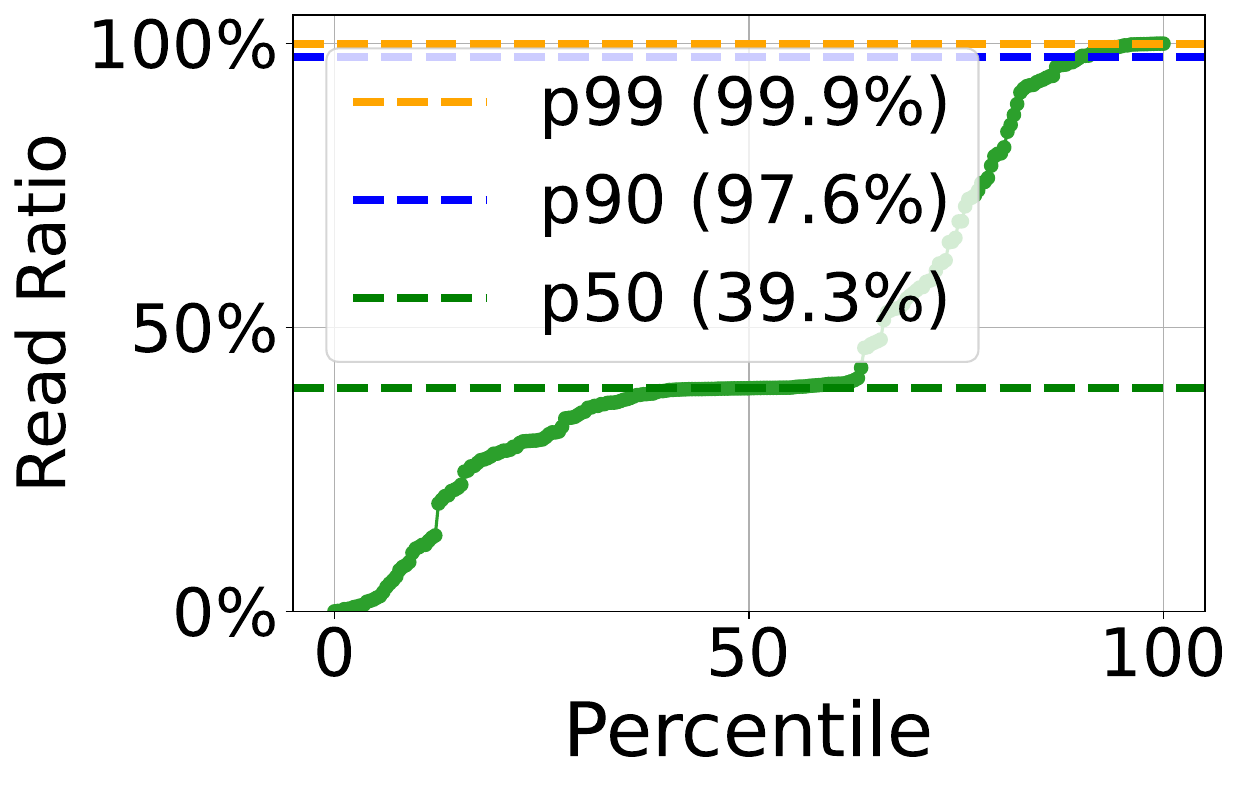}
        \caption{Read Ratio}
        \label{subfig:exp_read_ratio_perentile}
    \end{subfigure}
    \hfill
    \begin{subfigure}{0.23\textwidth}
        \includegraphics[width=\textwidth]{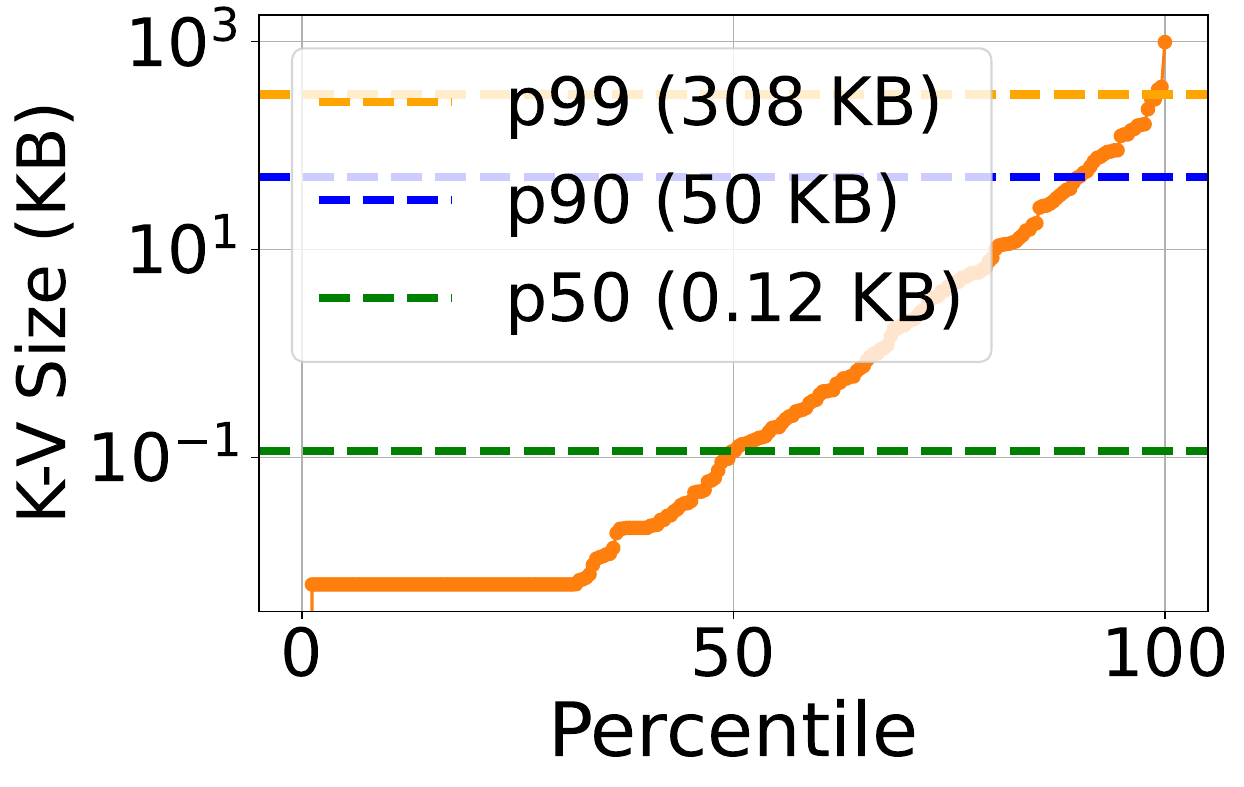}
        \caption{Average K-V Size}
        \label{subfig:exp_write_kv_size_perentile}
    \end{subfigure}
    \caption{Metric values across tenant percentiles.
    }
    \label{fig:exp_diversity_details}
\end{figure}

Figure~\ref{fig:exp_diversity_details} provides detailed metrics statistics. 
Figure~\ref{subfig:exp_latency_percentile} shows that all tenants on this resource pool experience latencies (P99) significantly below the SLA threshold (Service Level Agreement, red horizontal line). Exceeding SLA threshold indicates a failure to meet tenant demands. All tenants maintain latencies below 66.0\% of SLA, 90\% of tenants are under 24.0\% of SLA, and 50\% of tenants under 11.2\%. These low latencies demonstrate that ABase effectively supports diverse business needs and enhances performance stability, making it difficult for even significant traffic bursts to result in SLA violations. 
Figure~\ref{subfig:exp_cache_hit_ratio_perentile} shows the distribution of the cache hit ratio among tenants. 
Over 50\% of tenants have a cache hit ratio over 93.5\%, consistent with the low latency observed in most tenants (Figure~\ref{subfig:exp_latency_percentile}). 
Figure~\ref{subfig:exp_read_ratio_perentile} shows the distribution of read operation ratios among tenants: 50\% of ABase tenants have a read ratio of less than 39.3\% (write-heavy), while a significant proportion of tenants have a read ratio exceeding 50\% (read-heavy).
Finally, Figure~\ref{subfig:exp_write_kv_size_perentile} shows the distribution of the average key-value size among tenants. The median size is 0.12KB, with a few tenants having significantly larger sizes; on this resource pool, the 90th and 99th percentile key-value sizes are 50KB and 308KB, respectively.

\begin{figure}[tbp]
    \centering
    \begin{subfigure}[b]{0.23\textwidth}
        \centering
        \includegraphics[width=\textwidth]{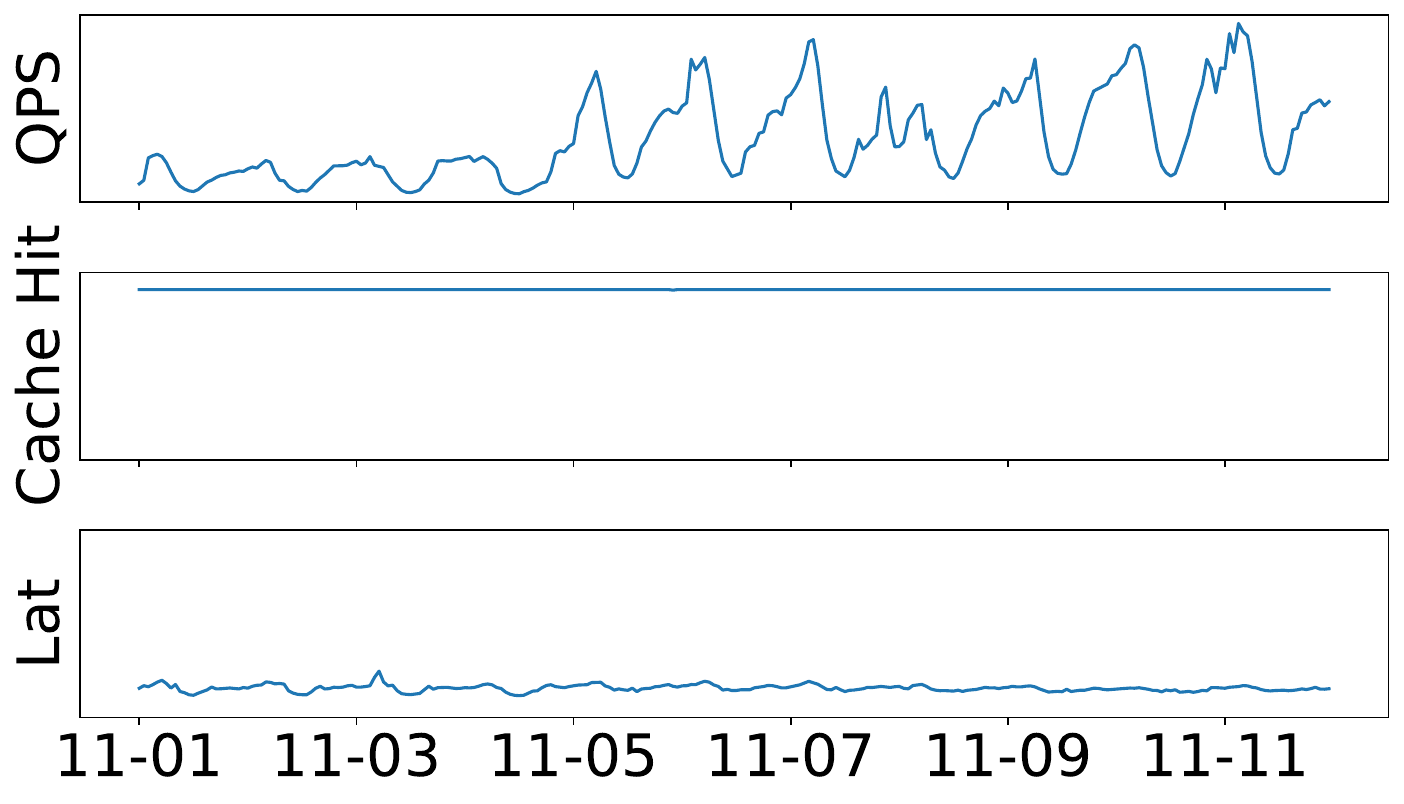}
        \caption{QPS increases, cache hit ratio remains stable.}
        \label{subfig:exp_qps_increase_cache_stable}
    \end{subfigure}
    \hfill
    \begin{subfigure}[b]{0.23\textwidth}
        \centering
        \includegraphics[width=\textwidth]{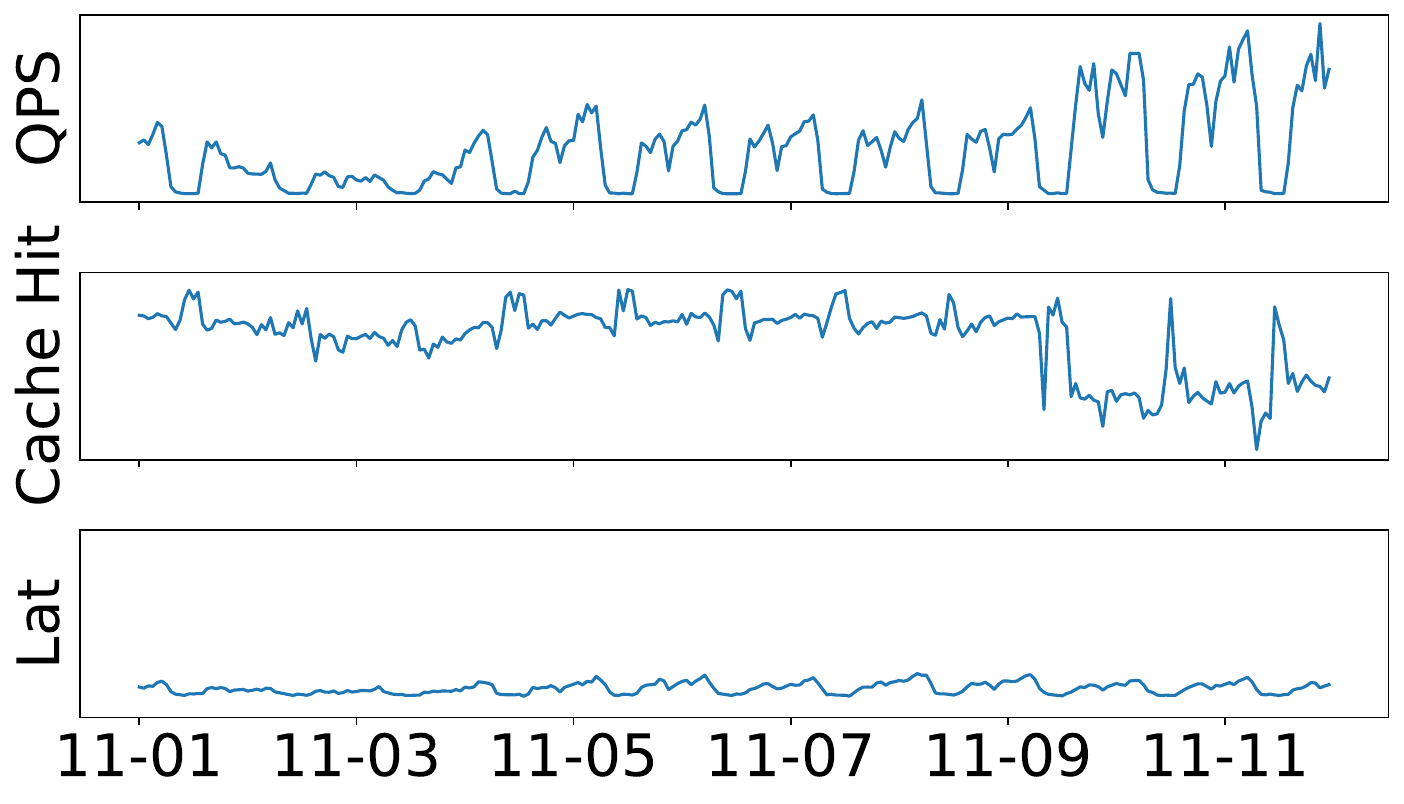}
        \caption{QPS increases, cache hit ratio decreases.}
        \label{subfig:exp_qps_increase_cache_drop}
    \end{subfigure}
    \begin{subfigure}[b]{0.23\textwidth}
        \centering
        \includegraphics[width=\textwidth]{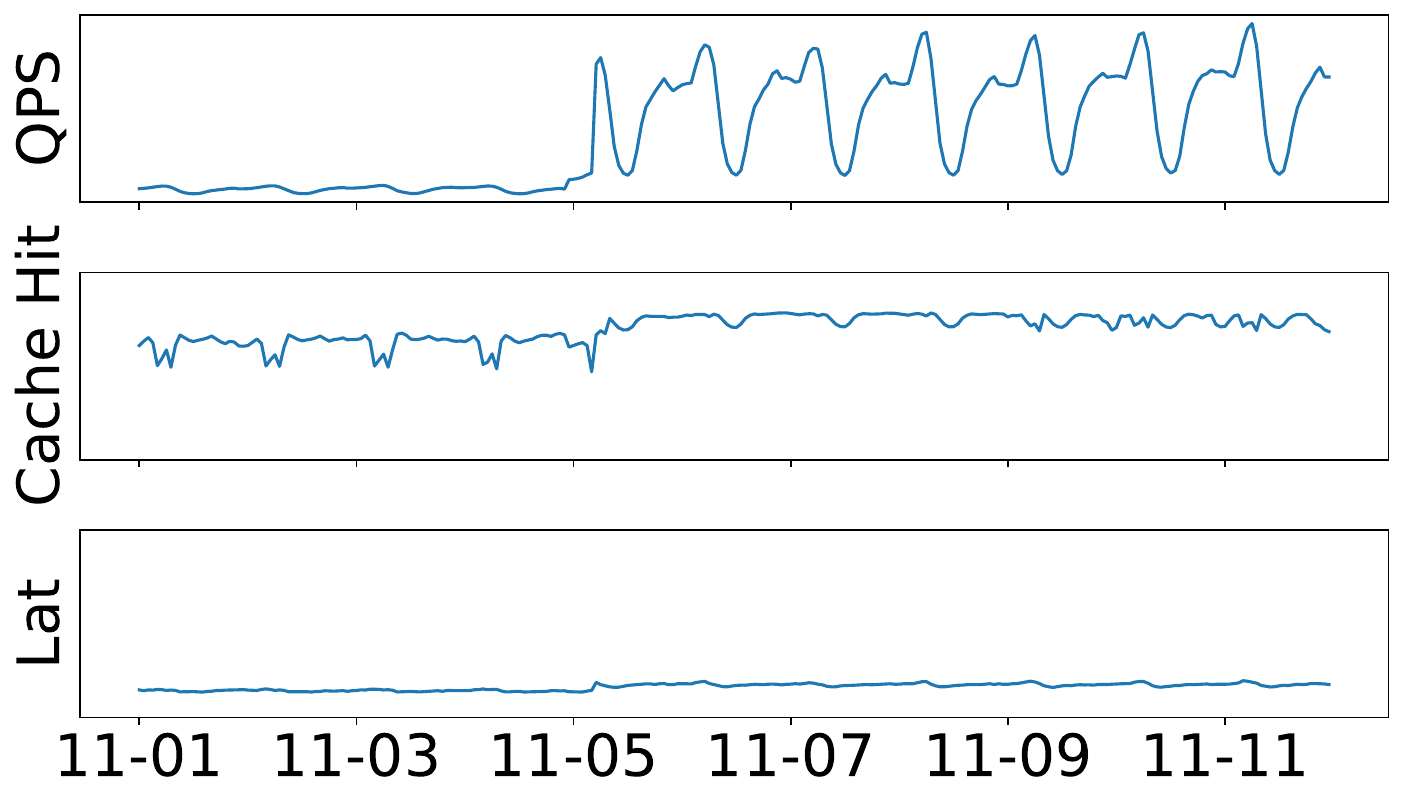}
        \caption{Both QPS and cache hit ratio increase.}
        \label{subfig:exp_qps_increase_cache_increase}
    \end{subfigure}
    \hfill
    \begin{subfigure}[b]{0.23\textwidth}
        \centering
        \includegraphics[width=\textwidth]{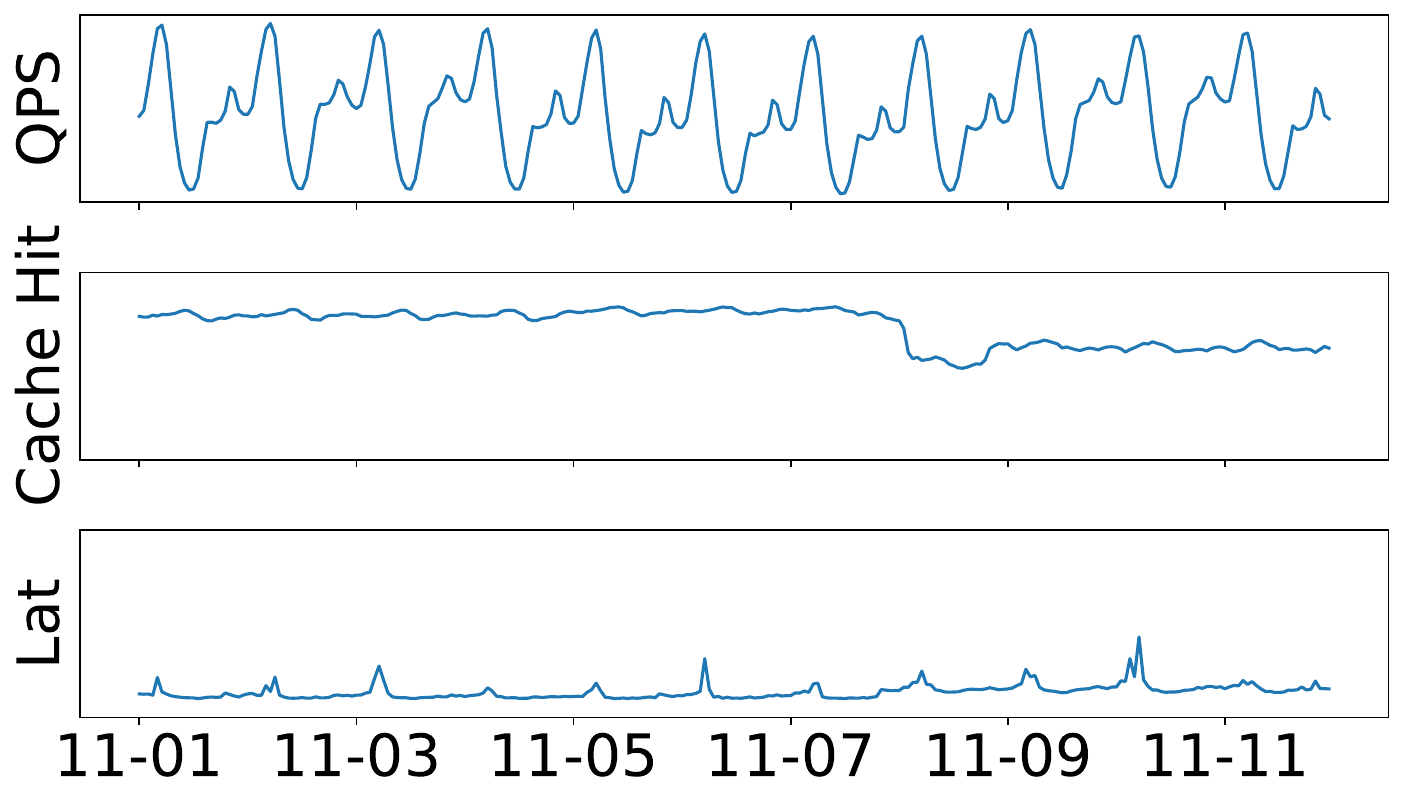}
        \caption{QPS remains stable, cache hit ratio decreases.}
        \label{subfig:exp_qps_stable_cache_decrease}
    \end{subfigure}
    \begin{subfigure}[b]{0.23\textwidth}
        \centering
        \includegraphics[width=\textwidth]{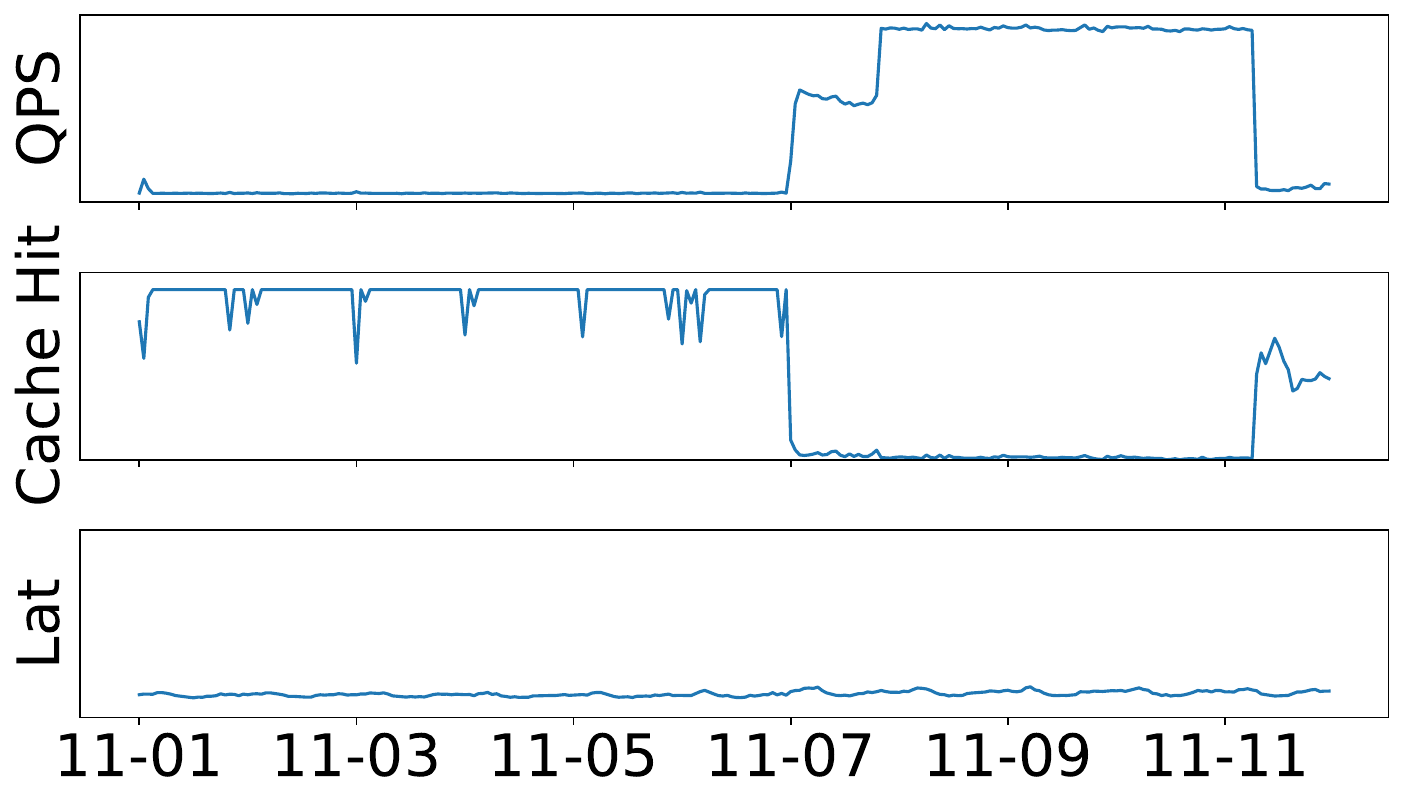}
        \caption{QPS increases shortly, cache hit ratio decreases.}
        \label{subfig:exp_qps_short_increase}
    \end{subfigure}
    \hfill
    \begin{subfigure}[b]{0.23\textwidth}
        \centering
        \includegraphics[width=\textwidth]{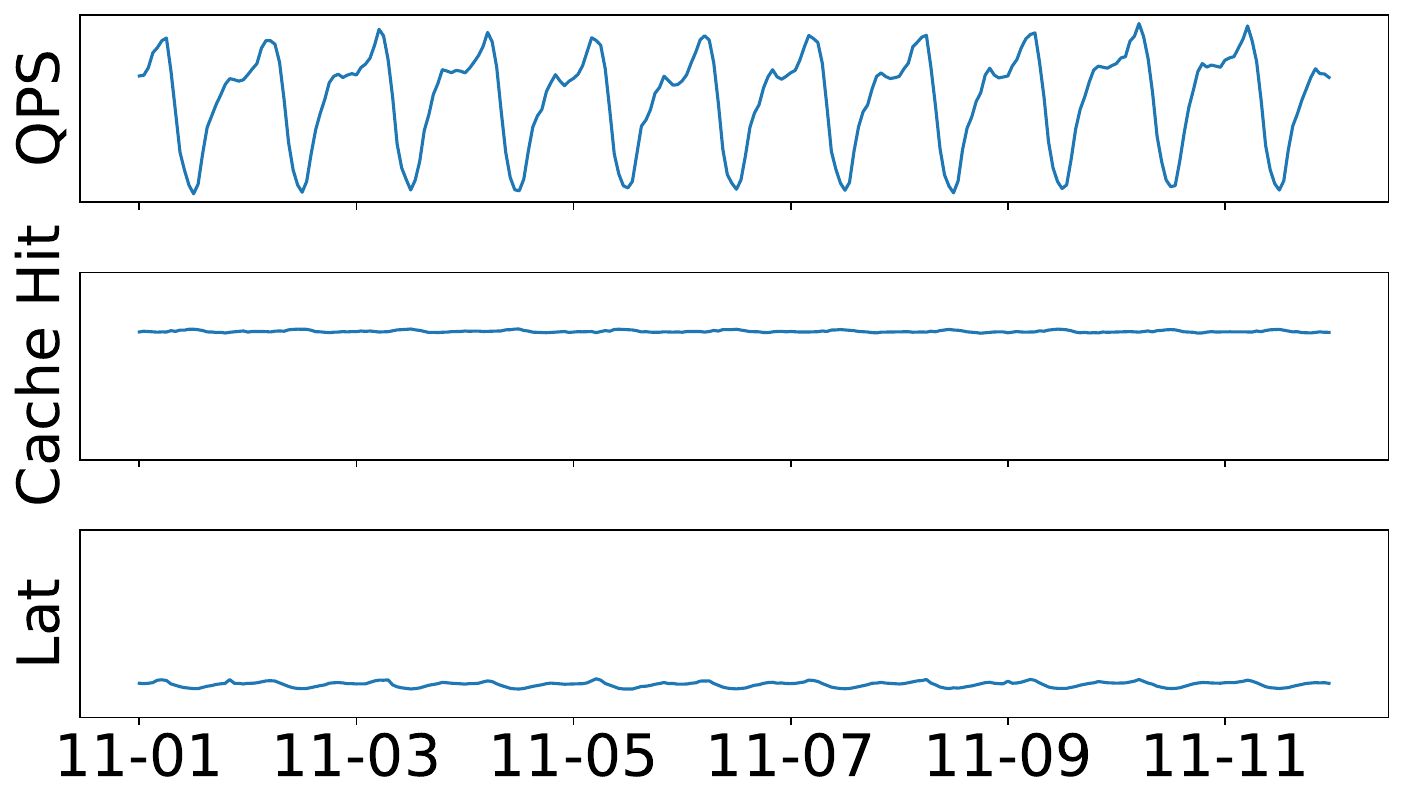}
        \caption{At resource pool scale: QPS and cache hit ratio remain stable.}
        \label{subfig:exp_cluster_stable}
    \end{subfigure}
    \caption{Tenant latency is stable amid workload fluctuations during the Double-11 Shopping Festival. Subfigures show QPS, cache hit ratio, and latency from top to bottom.}
    \label{fig:exp_dynamic_case}
\end{figure}

\noindent{\textbf{Dynamism Analysis:}} To demonstrate how effectively ABase handles the dynamic workloads at ByteDance, metrics were collected for tenants, including RU usage, cache miss ratio, and latency during the Double-11 Shopping Festival, a period characterized by intense E-Commerce activities that significantly alter the usual workload characteristics of many tenants. During the Double-11 period, more than 25\% of tenants in this resource pool exhibited significant increases in QPS or notable fluctuations in cache hit ratios. We illustrate some representative examples in Figure~\ref{fig:exp_diversity_details}. 
From Figure~\ref{subfig:exp_qps_increase_cache_stable} to Figure~\ref{subfig:exp_qps_increase_cache_increase}, all three tenants exhibited traffic increases, but their cache hit ratios varied. 
The cache hit ratio of Figure~\ref{subfig:exp_qps_increase_cache_stable} was virtually unaffected, remaining consistently at 100\%; 
in Figure~\ref{subfig:exp_qps_increase_cache_drop}, the cache hit ratio significantly decreased by over 20\% following an increase in tenant traffic, due to a broad distribution of requested keys leading to increased cache eviction.
Figure~\ref{subfig:exp_qps_increase_cache_increase} shows a 10\% increase in the cache hit ratio following a surge in tenant traffic, attributable to hot-key scenarios. In contrast to Figure~\ref{subfig:exp_qps_increase_cache_stable}, Figure~\ref{subfig:exp_qps_stable_cache_decrease} depicts a decrease of approximately 10\% in the cache hit ratio despite stable traffic levels. 
The tenant in Figure~\ref{subfig:exp_qps_short_increase} experienced a traffic peak lasting about 3 days, during which the cache hit ratio plummeted from 100\% to about 2\%. 

Despite the various workload changes among ABase's tenants, the latency for all tenants remained stable, still fully meeting the SLA requirements. This can be explained using Figure~\ref{subfig:exp_cluster_stable}, which shows changes in total traffic, average cache hit ratio, and average latency at the resource-pool level. Benefiting from ABase's multi-tenant design, the resource capacity of the resource pool far exceeds the variation in individual tenant demands, allowing tenants to share reserved resources and thus providing ample capacity to handle changes in tenant loads.
As a result, despite significant loads during the Double-11 shopping festival, overall pool traffic and cache hits remained stable.

\subsection{Performance Isolation}
\label{sub:isolation}
This section examines the effectiveness of the proxy quota, partition quota, and the dual-layer WFQ mechanism through ablation studies on synthetic workloads.

\begin{figure}[tbp]
    \centering
    \includegraphics[width=1\linewidth]{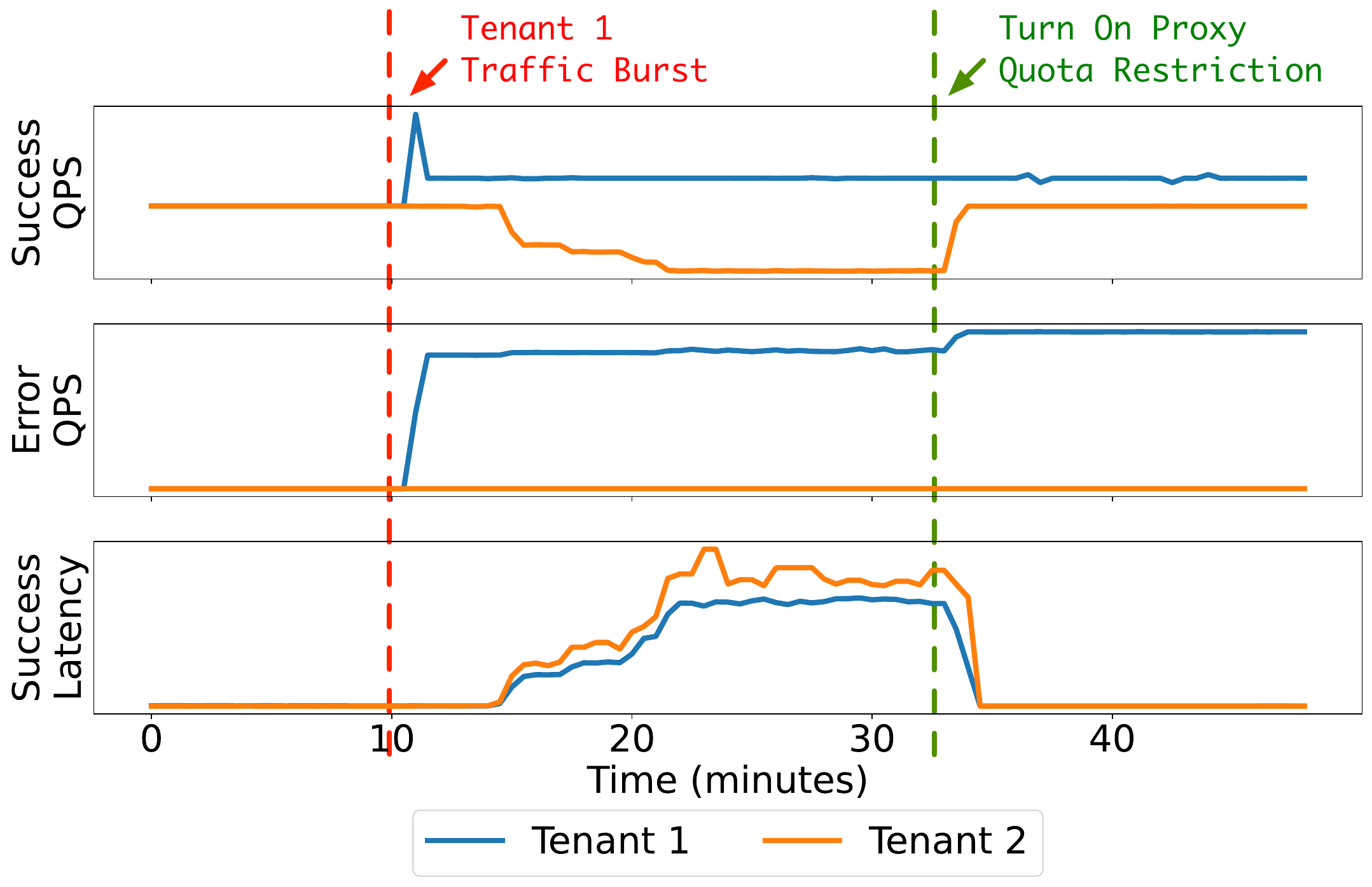}
    \caption{Effectiveness of proxy quota}
    \label{fig:isolation_exp_1_proxy_quota_miss}
\end{figure}

\textbf{Proxy Quota}.
As shown in Figure~\ref{fig:isolation_exp_1_proxy_quota_miss}, the experimental setup involved hosting partition replicas for two tenants on a single DataNode, with the proxy initially disabled.
Initially, both tenants experienced low traffic volumes, and all requests were processed successfully with minimal latency.
At the 10-minute mark, Tenant 1 initiated a traffic burst that significantly exceeded their assigned tenant quota (indicated by the red line).
In the absence of the proxy's interception, these requests overwhelmed the DataNode's request queue.
Tenant 1's success QPS reached the partition quota, and the requests exceeding this quota were returned as errors.
The DataNode expended considerable resources rejecting Tenant 1's excessive requests, which severely disrupted the processing of Tenant 2's legitimate requests.
Consequently, Tenant 2 was severely impacted by Tenant 1's burst, with their success QPS beginning to decline, nearly reaching zero.
At the 35-minute mark, upon activating Tenant 1's proxy (indicated by the green line),
the proxy efficiently intercepted traffic exceeding Tenant 1's tenant quota, enabling the DataNode to efficiently manage the remaining traffic.
Subsequently, latency levels for both tenants returned to low values, and Tenant 2's QPS recovered to the pre-burst levels.

\begin{figure}[tbp]
    \centering
    \includegraphics[width=1\linewidth]{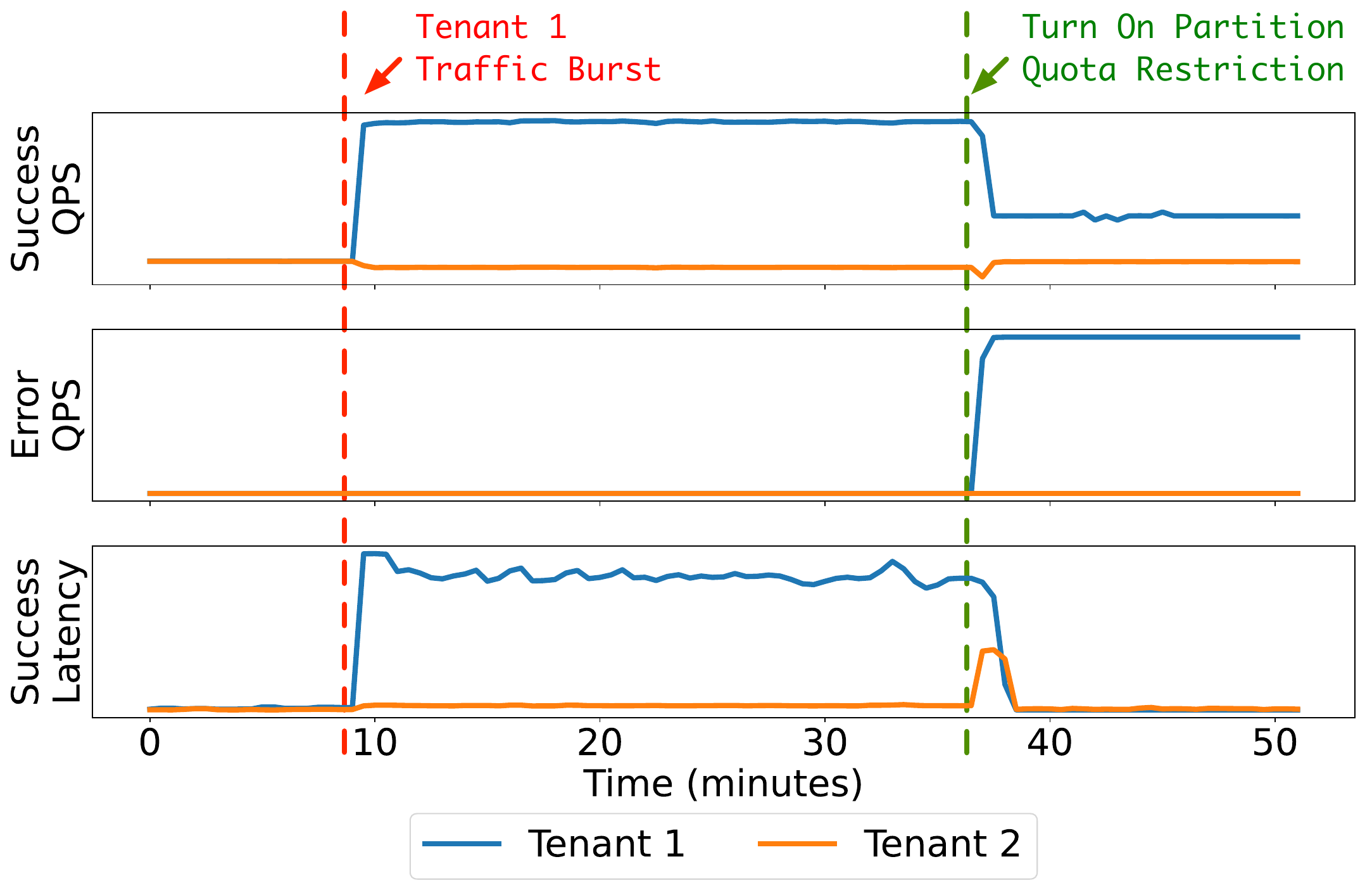}
    \caption{Effectiveness of partition quota and WFQ}
    \label{fig:isolation_exp2_partition_quota_and_wfq}
\end{figure}

\textbf{Partition Quota and Dual-Layer WFQ}. As shown in Figure~\ref{fig:isolation_exp2_partition_quota_and_wfq}, we conducted a simulation experiment to validate the efficacy of partition-level restrictions and the dual-layer WFQ mechanism. 
The setup, similar to a previous experiment, hosted two tenants' partition replicas on a single DataNode.
Initially, both tenants maintained normal QPS and latency levels under low traffic conditions, with the partition quota disabled.

At the 10-minute mark (indicated by the red line), we modeled a skewed partition traffic scenario for Tenant 1, directing a significant volume of traffic to Tenant 1's partition. Since the current traffic did not exceed the tenant quota, the proxy-level restriction did not reject any requests, resulting in zero error QPS for Tenant 1. Subsequently, the dual-layer WFQ mechanism was activated, aiming to ensure that the service capacity deployed on the DataNode's partition quota was proportional across tenants. Although Tenant 2's success QPS inevitably decreased by 25\%, the latency remained unaffected, indicating that the dual-layer WFQ mechanism preserved Tenant 2's isolation. However, for Tenant 1, the lack of DataNode limitations meant that ABase had to process all incoming requests, which led to a twenty-fold increase in latency, significantly degrading its quality of service.
At the 37-minute mark, we enabled the partition quota (indicated by the green line). Tenant 1's success QPS rapidly dropped to 3,000, matching the partition quota limit, and requests exceeding this threshold were rejected by the DataNode as error QPS. The success QPS for Tenant 2 also returned to its normal levels. Importantly, the latency for successful requests for both tenants was maintained at a low throughout the experiment.

\subsection{Elasticity}
\label{sub:elasticity}
\begin{figure}[tbp]
    \centering
    \begin{subfigure}[b]{0.23\textwidth}
        \centering
        \includegraphics[width=\textwidth]{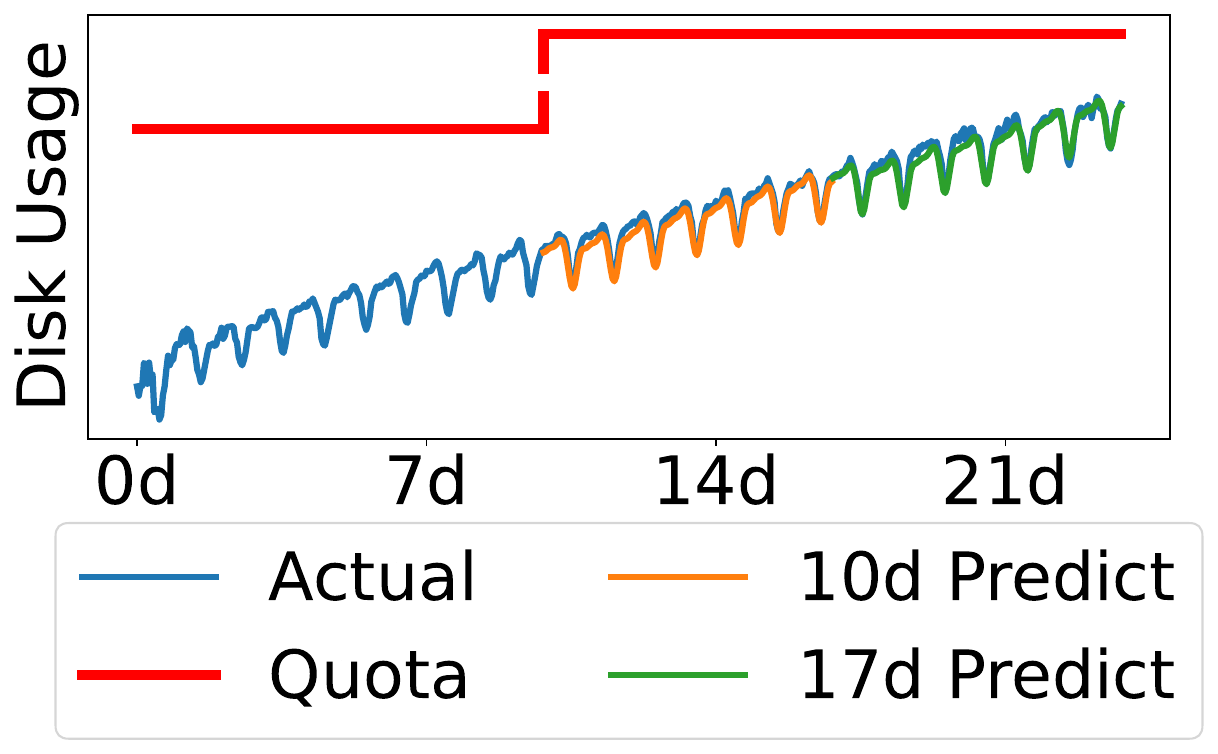}
        \caption{A scaling case}
        \label{fig:exp_scaling_exp1}
    \end{subfigure}
    \hfill
    \begin{subfigure}[b]{0.23\textwidth}
        \centering
        \includegraphics[width=\textwidth]{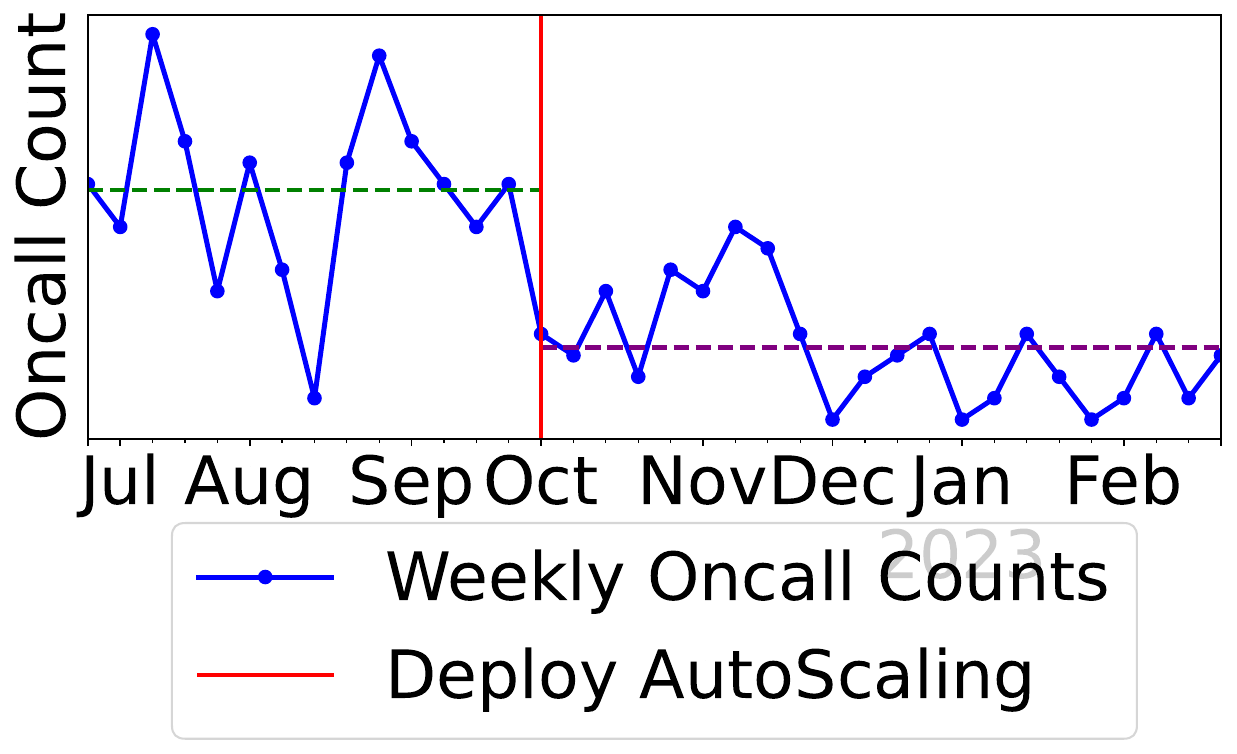}
        \caption{Oncall decrease}
        \label{fig:exp_scaling_exp2}
    \end{subfigure}
    \caption{Oncall (urgent contact) amount decreases by 65\%.}
    \label{fig:exp_scaling}
\end{figure}

This section shows the effectiveness of ABase's predictive scaling policy, using statistics from historical records. Figure~\ref{fig:exp_scaling_exp1} illustrates an online scaling example in the search business, where the disk usage (blue line) shows a 24-hour periodicity with an increasing trend. The tenant quota is depicted by the red line. On day 10, ABase predicted the usage would reach 85\% of the quota within a week (orange line), prompting a proactive quota increase to keep predicted usage below 65\%. This adjustment matched actual usage, as shown in Figure~\ref{fig:exp_scaling_exp1}, effectively preventing user throttling.

To demonstrate the business impact of the automatic scaling mechanism, we tracked the change in the number of upscaling oncalls 
(i.e. urgent contacts to technical support staff) 
over approximately six months before and after the deployment, as depicted in Figure~\ref{fig:exp_scaling_exp2}. Only up-scaling related oncalls are displayed. The occurrence of emergency oncalls likely indicates that users have experienced throttling, thus impacting the business. After deployment, the number of oncalls decreased by approximately 65\%, signifying a significant alleviation in user throttling.

\subsection{Resource Utilization}
\label{sub:resource_utilization}

\begin{figure}[htbp]
    \centering
    \begin{subfigure}[b]{0.23\textwidth}
        \centering
        \includegraphics[width=\textwidth]{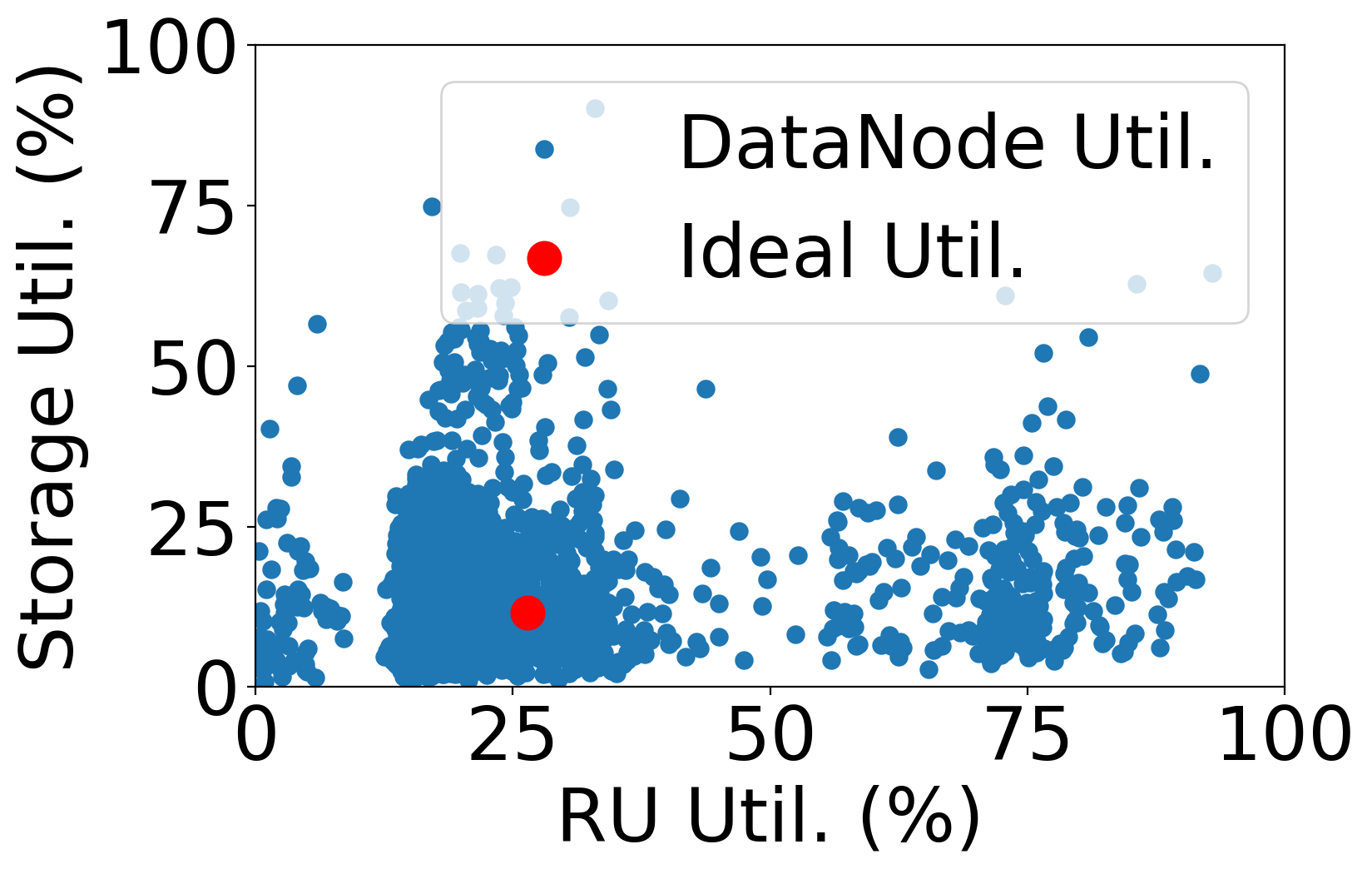}
        \caption{Before Rescheduling}
        \label{fig:rescheduling_exp2_before}
    \end{subfigure}
    \hfill
    \begin{subfigure}[b]{0.23\textwidth}
        \centering
        \includegraphics[width=\textwidth]{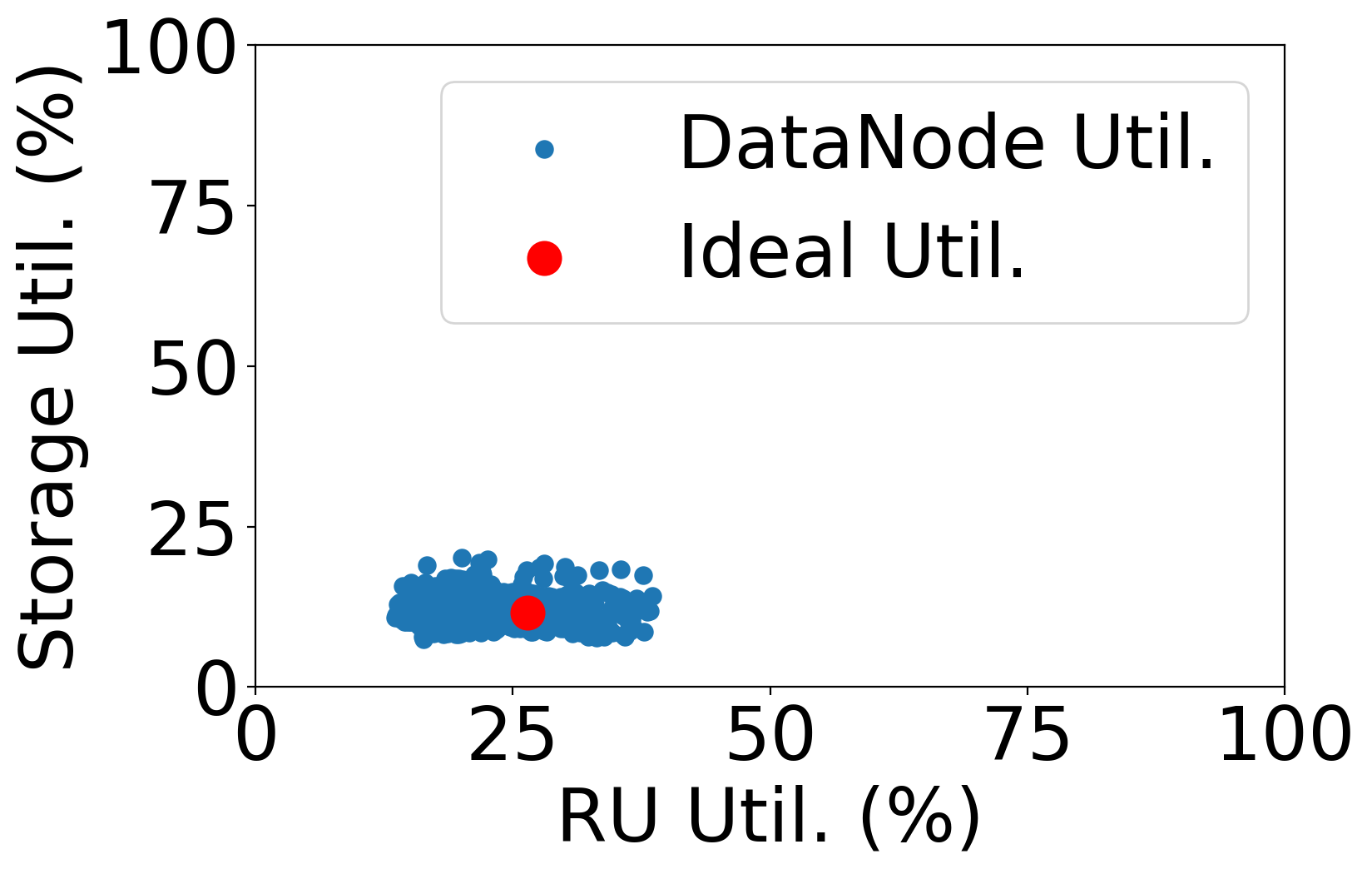}
        \caption{After Rescheduling}
        \label{fig:rescheduling_exp2_after}
    \end{subfigure}
    \caption{The resource utilization (Util.) for RU and Storage among DataNodes is improved through rescheduling.}
    \label{fig:offline_balance_graphs}
\end{figure}

To demonstrate the effectiveness of our rescheduling mechanisms, we first conducted offline experiments on a resource pool comprising 1000 DataNodes.
As shown in Figure~\ref{fig:rescheduling_exp2_before}, the original storage and RU utilization of the DataNodes were highly dispersed, indicating that the load on the DataNodes was extremely uneven, which limited the rapid scaling of tenants on them. Following the application of Algorithm~\ref{alg:rescheduling}, as shown in Figure~\ref{fig:rescheduling_exp2_after}, the load distribution across DataNodes was more balanced, with a 74.5\% reduction in the standard deviation of RU usage and an 84.8\% decrease in storage usage variance.

\begin{figure}[tbp]
    \centering
    \includegraphics[width=1\linewidth]{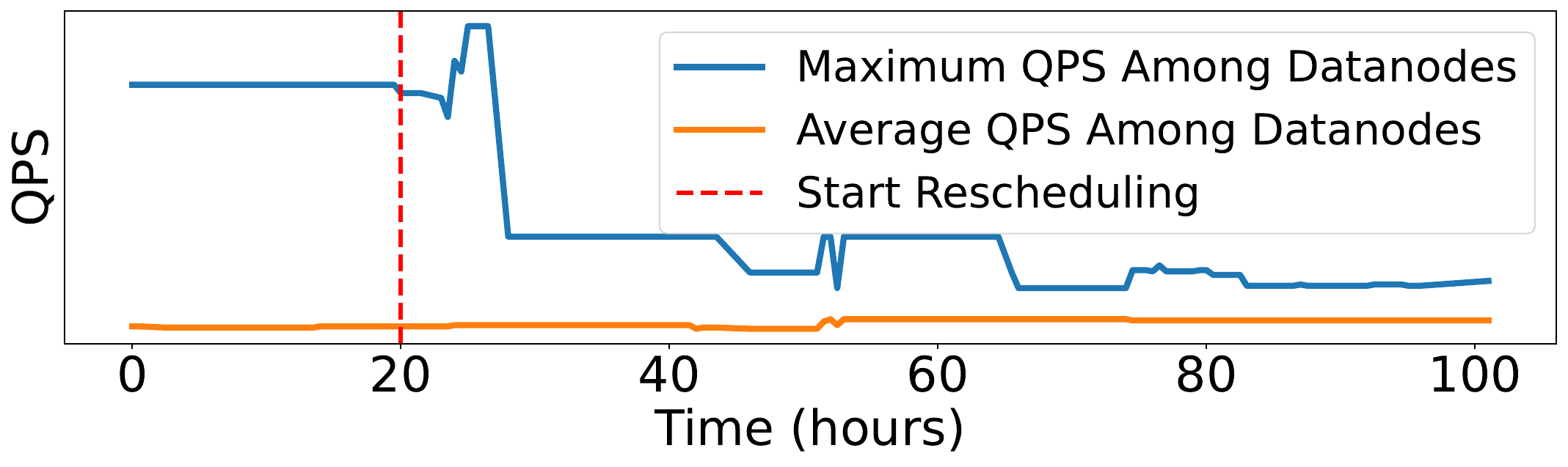}
    \caption{Rescheduling strategy reduces maximum QPS among DataNodes, indicating greater balance.}
    \label{fig:online_balance_graphs}
\end{figure}

This algorithm has been deployed in the online environment, executing once every 10 minutes. The changes in RU usage for a resource pool are illustrated in Figure~\ref{fig:online_balance_graphs}. 
Following the rescheduling algorithms, the maximum RU utilization among DataNodes increasingly converged towards the average RU utilization. Consequently, the proposed rescheduling algorithm effectively mitigates resource skewness, facilitating better resource utilization and reducing the risk associated with highly loaded DataNodes.

From the perspective of overall production statistics,
powered by the data rescheduling, ABase achieves higher resource utilization compared to the single-tenant ABase-Pre. The average utilization rates of CPU, Memory, and Disk for each machine in ABase-Pre were only 17\%, 52\%, and 27\%, respectively. After upgrading to ABase, these rates increased to 44\%, 63\%, and 46\%. This is because in the single-tenant design, the resources of low-utilization tenant cannot be reallocated to other tenants; moreover, as mentioned in Section~\ref{sec:architecture}, ABase-Pre must restrict the upper limit of resource utilization to tolerate single-node failures. Contrastively, the multi-tenant ABase eliminates machines with low utilization and enables resource pools to achieve higher utilization rates without sacrificing robustness.

\subsection{Cache Effectiveness}
\label{sub:exp_hot_key_caching}

\begin{table}[ht]
    \centering
    \caption{Benefit summary by proxy cache.}
    \resizebox{\columnwidth}{!}{%
        \begin{tabular}{lp{1cm}p{1cm}p{2cm}p{2cm}}
        \toprule
        Tenants  & \#Proxy & \#Group & Cache Hit Ratio & RU Saving \\
        \midrule
        Social Media 1 & 375     & 75      & $5\% \to 86\%$  & 85\%       \\
        Social Media 2 & 1626    & 32      & $5\% \to 67\%$  & 70\%       \\
        Social Media 3 & 11530   & 15      & $10\% \to 33\%$ & 38\%       \\
        E-Commerce 1   & 790     & 15      & $24\% \to 60\%$ & 61\%       \\ 
        E-Commerce 2   & 1511    & 15      & $24\% \to 60\%$ & 57\%       \\
        E-Commerce 3   & 4204    & 15      & $24\% \to 60\%$ & 79\%       \\
        
        \bottomrule
        \end{tabular}
    }
        
    \label{tab:exp_hot_key_caching_business}
\end{table}

We validated the effects of the proxy cache on six tenants within the Social Media and E-Commerce sectors. 
As shown in Table \ref{tab:exp_hot_key_caching_business}, the tenant Social Media 1 experiences extremely tight RU quotas during holiday periods, often resulting in throttling.
Despite having 375 proxies, the original cache hit ratio was only 5\%.
After activating the proxy cache and dividing the 375 proxies into 75 groups, this adjustment increased the cache hit ratio to 86\%, significantly reducing the underlying load and saving 85\% of RU for this tenant. Note that this change is very lightweight, solely altering the traffic routing proxy strategy. 
Similarly, for the remaining two Social Media tenants, the cache hit ratios improved by 62\% and 23\%, with RU savings of 70\% and 38\%, respectively. 
For the three E-commerce tenants, the cache hit ratios increased from 24\% to 60\%, with RU savings of 61\%, 57\%, and 79\% respectively.

\section{Lessons in Practice}
\label{sec:lessons}


\textbf{Resource Allocation}. We regulate the size of the resource pool to ensure that its idle resources exceed the quota of any single tenant. In practice, we ensure that the size of the resource pool is at least ten times the quota of any single tenant. Furthermore, at least 20\% of the resource pool consists of idle resources. This arrangement guarantees sufficient elasticity for any tenant while ensuring a controlled proportion of idle resources.

\textbf{Resource Isolation:} While increasing the scale of resource pools can enhance tenant elasticity when there is a significant proportion of idle resources, we recommend limiting the maximum number of tenants within a single resource pool and the maximum scale of each pool. Lessons learned from failures suggest that maintaining a moderate number of resource pools and tenants is crucial, which can avoid a large failure radius that could potentially lead to severe online incidents. Furthermore, given that the aggregate quota of resource pools should substantially exceed that of any individual tenant, we correspondingly regulate the maximum quota for each tenant.

\textbf{Handling Spiky Workloads}. To ensure rapid, second-level elastic scaling capabilities for tenants, we not only guarantee the reservation of idle resources at the entire resource pool level but also ensure a significant balance at the individual machine level. The idle resources are noticeably greater than any single tenant's quota at the same level, enabling each tenant to at least double their quota in the short term to accommodate sudden traffic changes.

\textbf{Auto-scaling Principles}. ABase approaches downscaling cautiously, prioritizing business stability. Overly aggressive downscaling might necessitate re-upscaling should business traffic rebound. Furthermore, even for tenants whose utilization has decreased but quota remains unscaled, this does not entail significant waste. 
A resource pool contains multiple tenants sharing idle resources. Some tenants' idle resources support others' burst traffic and growth, thereby maintaining a stable resource utilization rate.

\section{Related Works}
\label{sec:related_works}

\subsection{NoSQL Serverless Databases}
\label{sub:related_serverless}

Traditional NoSQL databases, such as Cassandra~\cite{lakshman2010cassandra}, have made substantial contributions to distributed database systems by emphasizing scalability, fault tolerance, and innovative consistency models, employing techniques such as sharding and replication. These systems excel in scenarios that require high write throughput and flexible schema designs for unstructured data. However, their architectures, originally designed for static resource allocation in single-tenant and on-premise environments, are deficient in native support for both elastic scaling and fine-grained performance isolation. This makes them less suited for cloud-native, multi-tenant serverless scenarios that demand dynamic resource provisioning and tenant-level SLA guarantees.

Multitenancy, an essential architectural approach for serverless databases, allows multiple tenants to share the same infrastructure, thereby enhancing scalability, flexibility, and cost-efficiency \cite{sellami2020mutlitenant}. However, this architecture poses significant challenges, including tenant isolation, load balancing, autoscaling, and issues related to hot keys \cite{kaur__SurveyDistributedData_2023}. 
DynamoDB~\cite{elhemali_ATC_DynamoDB_2022}, a pioneering serverless key-value NoSQL database, providing a scalable and predictably performant service, has set a benchmark for performance in distributed databases.
Although DynamoDB explores the requirement and necessity for traffic control and resource balancing in multi-tenant architectures, it does not disclose further technical details such as rescheduling algorithm and scaling policy.
As reported~\cite{elhemali_ATC_DynamoDB_2022}, DynamoDB supports trillions of API calls, peaking at 89.2 million QPS during the Amazon Prime Day shopping event. To support caching scenarios, DynamoDB introduces Amazon DynamoDB Accelerator (DAX)~\cite{amazon_dax}, supporting up to 10 nodes per tenant and millions of QPS. 
Microsoft CosmosDB~\cite{azure_cosmos_db} offers a fully managed serverless experience but imposes a capacity limit of 1 million request units per database, limiting its ability to handle large-scale workloads. Firestore~\cite{kesavan_ICDE_Firestore_2023} is tailored to enhance usability for web and mobile developers, offering real-time data synchronization and scalable development within the Firebase ecosystem.


\subsection{Predictive AutoScaling}
Autoscaling in cloud systems has drawn significant attention, with notable contributions from Qu et al.~\cite{qu2018auto}, Barnawi et al.~\cite{barnawi2020views}, and Lorido-Botran et al.~\cite{lorido2014review}.
These solutions are now extensively implemented across a variety of infrastructure services, such as databases~\cite{taft2018p, lolos2017adaptive, jindal2019peregrine} and microservices~\cite{yu2020microscaler, abdullah2020burst, bauer2019chamulteon}. Autoscaling is typically categorized by scaling direction into horizontal~\cite{zhou_AAAI_AHPA_2023} and vertical~\cite{rzadca_EuroSys_Autopilot_2020} types, as well as by timing into reactive and proactive types~\cite{qu2018auto}. This paper concentrates on predictive scaling.

Workloads exhibiting regular periods have been shown to significantly benefit from proactive strategies as demonstrated by Higginson et al.~\cite{higginson_SIGMOD20_DatabaseWorkloadCapacity_2020} and Cortez et al.~\cite{cortez2017resource}. However, the diversity of periods and trends introduces substantial forecasting challenges. To address these, Qin et al. have proposed a collection of robust decomposition methods~\cite{qian_ICDE_RobustScaler_2022, wen2021robustperiod, wen2020fast}. Moreover, integrating multiple prediction models has proven effective in handling complex workload patterns in industrial applications. For instance, Seagull~\cite{poppe_VLDB_Seagull_2020} classifies Microsoft Azure services into daily/weekly and stable/short-lived categories based on user activity, applying tailored prediction models for each. Kim et al. introduce a cloud workload prediction framework that incorporates multiple predictors~\cite{kim__ForecastingCloudApplication_2022}. Hu et al. describe a framework that integrates five distinct prediction models for effective virtual machine provisioning~\cite{hu__AutoscalingPredictionModels_2016}.

\subsection{Resource Scheduling}
Resource scheduling in cloud computing has been extensively studied in recent years. For example, Eigen~\cite{li2023eigen} introduces a hierarchical resource management system, along with three heuristic-based resource optimization algorithms aimed at enhancing the resource allocation ratio without compromising resource availability. K{\"o}niget al.~\cite{konig2023solver} propose a method that combines mathematical modeling with solvers to address the tenant placement problem in a Database-as-a-Service cluster, with a focus on minimizing the probability of failovers. Chen et al.~\cite{Chen_ICDE_RASA_2024} develop a method based on graph partitioning and solver-based algorithms to address resource allocation with service affinity in large-scale cloud environments. RAS~\cite{newell2021ras} employs Mixed-Integer Programming (MIP) to formulate the capacity reservation challenge for large-scale clusters. To adhere to the Service Level Objective (SLO) of achieving a solution within one hour, multi-phase solving techniques and variable aggregation methods are utilized.

\section{Conclusion}
\label{sec:conclusion}

This paper introduces ABase, a multi-tenant NoSQL serverless database developed at ByteDance. We analyze the diverse and dynamic nature of workloads, summarize the challenges, and detail our contributions. Firstly, ABase introduces dual-layer caching to support high-speed caching scenarios, alongside a cache-aware isolation mechanism to address the impacts of cache hits on resource consumption estimates. Secondly, ABase has developed a predictive autoscaling policy to dynamically adjust resources in alignment with actual demand. Additionally, ABase proposes a limited fan-out hash strategy to mitigate impacts from hot key pressure or a decline in cache hits.
Finally, ABase introduces a multi-resource rescheduling algorithm to balance resource utilization across data nodes. ABase has supported a workload with a peak QPS of over 13 billion and storage over 1 EB, and the experiments and production analysis have validated the effectiveness of ABase's innovations.

\bibliographystyle{ACM-Reference-Format}
\balance
\bibliography{abase}

\end{document}